\documentclass[journal,10pt]{IEEEtran}
\usepackage{graphicx}
\usepackage{subfigure}
\usepackage{float}
\usepackage{subfloat}
\usepackage{amsmath}
\usepackage{amssymb}
\usepackage{amsthm}
\usepackage{epstopdf}
\usepackage{cases}
\usepackage{color}
\usepackage{makecell}
\usepackage{cite}
\usepackage{algorithmic}
\usepackage[ruled]{algorithm2e}

\newtheorem{proposition}{Proposition}

\usepackage[normalem]{ulem}

\ifCLASSINFOpdf
\else
\fi
\usepackage{caption}
\usepackage{mathtools}

\begin{document}

%
\title{Joint Localization and Communication Enhancement in Uplink Integrated Sensing and Communications System with Clock Asynchronism}
%
%
%

\author{Xu Chen~\footnotemark{*},~\IEEEmembership{Member,~IEEE,}
	XinXin He~\footnotemark{*},~\IEEEmembership{Member,~IEEE,}
	Zhiyong Feng,~\IEEEmembership{Senior Member,~IEEE,}
	Zhiqing Wei,~\IEEEmembership{Member,~IEEE,}
	Qixun Zhang,~\IEEEmembership{Member,~IEEE,}
	Xin Yuan,~\IEEEmembership{Member,~IEEE,}
	and Ping Zhang,~\IEEEmembership{Fellow,~IEEE}
	\thanks{This work is supported by the National Key Research and Development Program of	China under Grants \{2020YFA0711300, 2020YFA0711302, and 2020YFA0711303\}.
	}
	\thanks{*These authors contributed equally to this work and should be regarded as co-first authors.}
	\thanks{Xu Chen, Xinxin He, Z. Feng, Zhiqing Wei, and Qixun Zhang are with School of Information and communication Engineering, Beijing University of Posts and Telecommunications, Beijing 100876, P. R. China (Email:\{chenxu96330, hxx\_9000, fengzy, weizhiqing, zhangqixun\}@bupt.edu.cn).}
	\thanks{X. Yuan is with the University of Technology Sydney, Ultimo, NSW 2007, Australia (email: xin.yuan@ieee.org).}
	\thanks{Ping Zhang is with State Key Laboratory of Networking and Switching Technology, Beijing University of Posts and Telecommunications, Beijing 100876, P. R. China (Email: pzhang@bupt.edu.cn).}
	\thanks{Corresponding author: Zhiyong Feng}
}

%
%

\markboth{}%
{Shell \MakeLowercase{\textit{et al.}}: Bare Demo of IEEEtran.cls for IEEE Journals}
%


\maketitle

\newcounter{mytempeqncnt}
\setcounter{mytempeqncnt}{\value{equation}}
\begin{abstract}
	In this paper, we propose a joint single-base localization and communication enhancement scheme for the uplink (UL) integrated sensing and communications (ISAC) system with asynchronism, which can achieve accurate single-base localization of user equipment (UE) and significantly improve the communication reliability despite the existence of timing offset (TO) due to the clock asynchronism between UE and base station (BS). Our proposed scheme integrates the CSI enhancement into the multiple signal classification (MUSIC)-based AoA estimation and thus imposes no extra complexity on the ISAC system. We further exploit a MUSIC-based range estimation method and prove that it can suppress the time-varying TO-related phase terms. Exploiting the AoA and range estimation of UE, we can estimate the location of UE. Finally, we propose a joint CSI and data signals-based localization scheme that can coherently exploit the data and the CSI signals to improve the AoA and range estimation, which further enhances the single-base localization of UE. The extensive simulation results show that the enhanced CSI can achieve equivalent bit error rate performance to the minimum mean square error (MMSE) CSI estimator. The proposed joint CSI and data signals-based localization scheme can achieve decimeter-level localization accuracy despite the existing clock asynchronism and improve the localization mean square error (MSE) by about 8 dB compared with the maximum likelihood (ML)-based benchmark method. 

\end{abstract}


%
\IEEEpeerreviewmaketitle

\section{Introduction}
%
%
%
%
\subsection{Backgrounds and Motivations}
With the rapid development of the sixth-generation (6G) intelligent machine-type applications, such as intelligent vehicular networks, smart factories, and smart cities, both communication and sensing are indispensable for the flexible operation of autonomous machines~\cite{Feng2021JCSC}. Integrated sensing and communications (ISAC), also known as joint communication and sensing (JCAS), is a promising technique to solve the spectrum congestion problem due to the proliferation of wireless communication and sensing devices~\cite{liu2020joint, Chen2021CDOFDM}. ISAC can share the same transceivers and spectrum to achieve both sensing and communication functions using the same transmitted signals~\cite{Feng2021JCSC, ZhangOverviewJCS, Yuan2021}.  

ISAC achieved by using uplink (UL) signals, i.e., UL ISAC, has become a promising approach to deploying ISAC techniques in the cellular communication systems~\cite{Zhang2022ISAC}, since it does not require full-duplex (FD) operations as in the downlink (DL) ISAC~\cite{IBFDJCR}. One example of UL ISAC in communication systems is to conduct sensing at the base station (BS) using the UL communication signals from user equipment (UE). Consequently, UL ISAC can be implemented with minimal adjustments to the network infrastructure. 
Nonetheless, the disparity in clock synchronization between the BS and UE introduces a fluctuating timing offset (TO), posing a substantial obstacle to achieving precise range and location estimations in UL ISAC. Therefore, mitigating the sensing ambiguity from clock asynchronism is one of the biggest challenges in UL ISAC.

In this paper, we focus on the single-base localization of UE in UL ISAC, which does not require solving the complex clock synchronization issues among multiple BSs as conducted in the multi-base cooperative localization~\cite{2021Girimlocalization}. Moreover, as illustrated in the early works~\cite{Chen2023KF}, the angle-of-arrival (AoA) estimation in UL ISAC is not prominently affected by the clock asynchronism. Therefore, using the AoA and range estimation of a single BS to localize the UE can avoid the partial influence of sensing ambiguity. Furthermore, we consider enhancing the channel state information (CSI) estimation in the process of AoA estimation in UL ISAC to improve communication reliability.

\subsection{Related Works}
In this subsection, we review the existing works related to the UL ISAC localization and CSI estimation.

For a long time, cooperation among multiple BSs and iterative location estimation using multiple packets has been prevalent to deal with the range and location offset due to TO. In~\cite{2019YuanLocalization}, the authors proposed an expectation-maximization-based cooperative localization method. This method requires numerous receivers and more than 20 iterations for expectation maximization, which results in a high implementation complexity. In~\cite{2021Girimlocalization}, the authors proposed a cooperative beamforming and power allocation scheme with multiple BSs to maximize the maximum rate and positioning error bounds. However, this paper does not present the implementable signal processing method for achieving the localization performance bounds and ignores the influence of TO. 

Recently, new techniques based on the feature of multiple antennas have been proposed to address the clock asynchronism problem. 
In~\cite{Qian2018Widar2}, the authors proposed to use the cross-antenna cross-correlation (CACC) method to track humans passively with a single WiFi link by exploiting the cross-correlation among all the pairs of antennas.
In~\cite{Nizhitong2021}, the authors proposed a UL ISAC method for perceptive mobile networks, allowing a static UE and BS to form a bi-static system to sense the environment by utilizing the CACC method. However, this method only works under the assumption that the accurate location of UE is known in advance and has to solve the challenging image target problems, which means it is unsuitable for localizing UEs. In~\cite{ZhangDaqing2019,ZhangDaqing2020,Li2022CSIsensing}, the authors proposed to use the cross-antenna signal ratio (CASR) method to estimate the Doppler frequency exploiting the CSI ratio between each two antennas. However, this method works only when the scatterers are static except for the target of interest to maintain the linearity of the sensing parameter estimation problem based on the CSI ratio. 

On the other hand, the CSI estimators and enhancers are also promising research areas in ISAC. The least-square (LS) method is widely used to estimate CSI for communication demodulation and sensing parameter estimation~\cite{Sturm2011Waveform, Zhang2019JCRS} due to its low complexity. In~\cite{2023XuJCAS}, the authors proposed to use the sensing parameter estimates to enhance the CSI estimation accuracy for improving communication reliability. In~\cite{2023ChenKalman}, the authors proposed a Kalman filter (KF)-based communication CSI enhancer by exploiting the AoA estimates to suppress the random noise terms in LS CSI to improve the communication performance. However, these CSI enhancers require adding filtering procedures into the original ISAC signal processing, which imposes extra complexity. 

\subsection{Contributions}
In this paper, we propose a novel joint single-base localization and communication enhancement scheme for UL ISAC. We first propose a joint AoA estimation and CSI enhancement method, which integrates the CSI enhancement into the multiple signal classification (MUSIC)-based AoA estimation procedures. Unlike the existing CSI enhancers, this method is completed in the AoA estimation and imposes no extra complexity on the ISAC system. Subsequently, we further reveal that the MUSIC-based range estimation method can suppress the TO-related noise-like phase terms in CSI estimation, which can reduce the range ambiguity. Using the AoA and range estimation, we can estimate the location of UE with spherical coordinates. Finally, we propose a novel joint CSI and data signals-based localization scheme that uses the combined demodulated data and CSI signals to improve the sensing performance. 

The main contributions of this paper are summarized as follows. 
\begin{itemize}
	
	\item[1.] We propose a joint AoA estimation and CSI enhancement method that uses the eigenvalue decomposition procedure in the MUSIC-based AoA estimation to suppress the noise terms in the initial CSI estimation obtained by LS estimator. This method requires no additional filters to enhance the CSI estimation and imposes no extra complexity to the ISAC system. The simulation results show that the proposed joint AoA estimation and CSI enhancement method can achieve the same bit error rate (BER) performance as the minimum mean square error (MMSE) estimator.
	
	\item[2.] We reveal that the MUSIC-based range estimation method can suppress the TO-related time-varying noise terms by exploiting the time-domain averaging effects on multiple packets in the eigenvalue decomposition procedures. Moreover, the range estimation of UE is not affected by the noise and offset terms in frequency and antenna domains since the frequency and antenna-domain noise is coherently absorbed in the eigenvalue decomposition procedures.
	
	\item[3.] We propose a novel joint CSI and data signals-based localization scheme that can flexibly combine the CSI and demodulated data signals to improve the AoA and range estimation, which naturally improves the localization performance. We prove that the MUSIC-based AoA and range estimation can coherently utilize the data signals to improve the AoA and range estimation. 
\end{itemize}

We provide extensive simulation results, validating the proposed joint single-base localization and communication enhancement scheme. The localization mean square error (MSE) of the proposed scheme is shown to be about 8 dB lower than the maximum-likelihood (ML)-based benchmark scheme.

\subsection{Outline of This Paper}

The remaining parts of this paper are organized as follows. 
In section \ref{sec:system-model}, we describe the system model of the UL JCAS scheme. 
Section \ref{sec:JCAS_sensing} proposes the joint single-base localization and CSI enhancement scheme.
Section \ref{sec:Joint_CSI_data_localization} proposes the joint CSI and data signals-based localization scheme and analyzes its complexity.
In section \ref{sec:Simulation}, the simulation results are presented to validate the proposed scheme. 
Section \ref{sec:conclusion} concludes this paper.

\begin{figure}[!t]
	\centering
	\includegraphics[width=0.30\textheight]{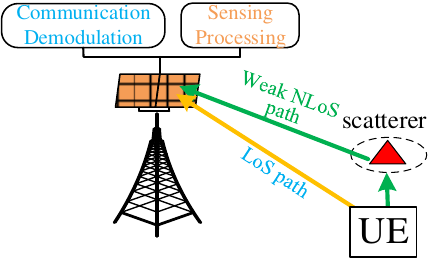}%
	\DeclareGraphicsExtensions.
	\caption{The UL ISAC scenario.}
	\label{fig: Uplink JCS Model}
\end{figure}

\textbf{Notations}: Bold uppercase letters denote matrices (e.g., $\textbf{M}$); bold lowercase letters denote column vectors (e.g., $\textbf{v}$); scalars are denoted by normal font (e.g., $\gamma$); the entries of vectors or matrices are referred to with square brackets, for instance, the $q$th entry of vector $\textbf{v}$ is $[\textbf{v}]_{q}$, and the entry of the matrix $\textbf{M}$ at the $m$th row and $q$th column is ${[\textbf{M}]_{n,m}}$; ${{\bf{U}}_s} = {\left[ {\bf{U}} \right]_{:,{N_1}:{N_2}}}$ means the matrices sliced from the $N_1$th to the $N_2$th columns of $\bf U$; $\left(\cdot\right)^H$, $\left(\cdot\right)^{*}$ and $\left(\cdot\right)^T$ denote Hermitian transpose, complex conjugate and transpose, respectively; ${\rm{vec}}(\cdot)$ is the operation to vectorize a matrix; ${\bf M}_1 \in \mathbb{C}^{M\times N}$ and ${\bf M}_2 \in \mathbb{R}^{M\times N}$ are ${M\times N}$ complex-value and real-value matrices, respectively; ${\left\| {\mathbf{v}}  \right\|_l}$ represents the ${ \ell}$-norm of ${\mathbf{v}}$, and $\ell_2$-norm is considered in this paper; for two given matrices ${\bf{S}}_1$ and ${\bf{S}}_2$,  ${ {[ {{v_{p,q}}} ]} |_{(p,q) \in {\bf{S}}_1 \times {\bf{S}}_2}}$ denotes the vector stacked by values ${v_{p,q}}$ satisfying $p\in{\bf{S}}_1$ and $q\in{\bf{S}}_2$; and $v \sim \mathcal{CN}(m,\sigma^2)$ means $v$ follows a circular symmetric complex Gaussian (CSCG) distribution with mean value $m$ and variance $\sigma^2$.

\begin{figure*}[!t]
	\centering
	\includegraphics[width=0.60\textheight]{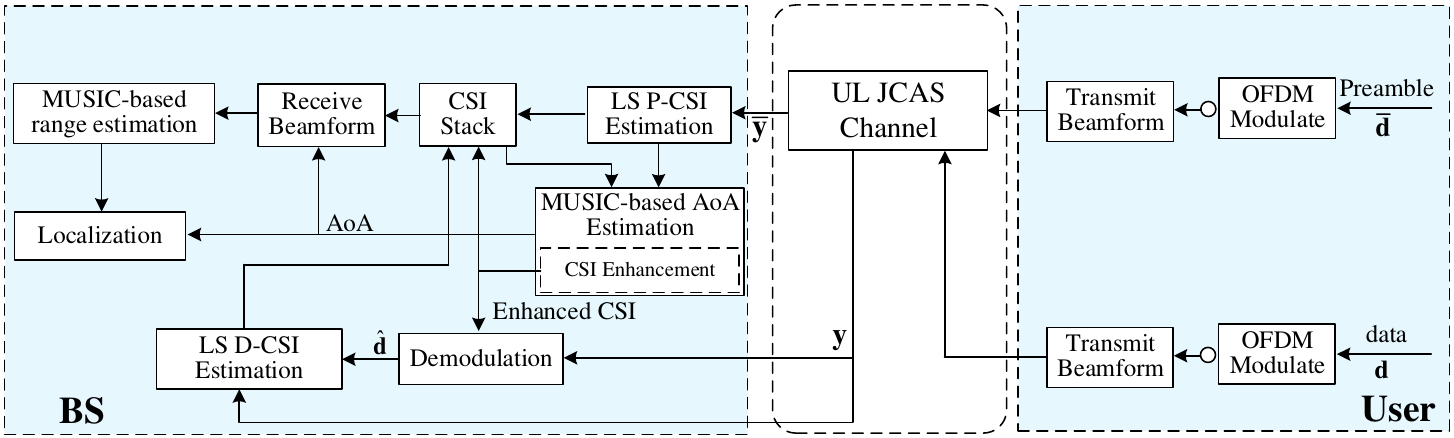}%
	\DeclareGraphicsExtensions.
	\caption{The diagram for joint single-based localization and communication enhancement.}
	\label{fig: UL communication}
\end{figure*}

\section{System Model}\label{sec:system-model}
This section presents the UL ISAC system setup, channel models, and received signal models to provide fundamentals for UL ISAC signal processing. 

\subsection{UL ISAC System Setup} \label{subsec:The Uplink JCS Model}
We consider a UL ISAC system where the BS and UE use uniform plane arrays (UPAs), as shown in Fig.~\ref{fig: Uplink JCS Model}. In the UL preamble period, UE transmits the pilot signals to BS, and BS uses the received training sequences for CSI estimation. In the UL data period, UE transmits the data signals to BS, and BS demodulates the UL data signals of UE using the estimated CSI. 

In this paper, the CSI estimation and data symbols are both employed to enhance the estimation of the AoA and range to improve the localization performance. We assume that the pilot sequences for channel estimation are transmitted at an equal interval for simplicity of presentation. Each training sequence can be part of a typical packet in, e.g., WiFi systems, or transmitted in timeslot regularly, in, e.g., 5G mobile networks. There is one channel estimate during each packet. We also assume that within a coherent processing interval, $M_s$ CSI estimates are obtained at an interval of $T_s^p$. The key idea in this paper can be extended to more general cases of non-uniform intervals. 

Orthogonal frequency division multiplexing (OFDM) signal is utilized in the ISAC system. The key parameters for the OFDM signal are denoted as follows. ${P_t^U}$ is the transmit power, ${N_c}$ is the number of subcarriers occupied by UE in each OFDM symbol; $f_c$ is the carrier frequency, $T_s$ is the time duration of each symbol, and $\Delta f$ is the subcarrier interval. Since there is one CSI estimate in each packet, $T_s^p = P_s T_s$, where $P_s$ is the number of OFDM symbols in each packet.


\subsection{UPA Model} \label{subsec:UPA model}
The size of UPA is ${P} \times {Q}$, and we use $d_a$ to denote the uniform interval between the neighboring antenna elements. The AoA for receiving or the angle-of-departure (AoD) for transmitting the $k$th far-field signal is ${{\bf{p}}_k} = {( {{\varphi _k},{\theta _k}} )^T}$, where ${\varphi _k}$ and ${\theta _k}$ are the azimuth and elevation angles, respectively. The phase difference between the ($p$,$q$)th and the reference antenna elements is~\cite{Chen2020}
\begin{equation}\label{equ:phase_difference}
	{a_{p,q}}\left( {{{\bf{p}}_k}} \right) \! =\! \exp [ { - j\frac{{2\pi }}{\lambda }{d_a}( {p\cos {\varphi _k}\sin {\theta _k} \!+\! q\sin {\varphi _k}\sin {\theta _k}} )} ],
\end{equation}
where $\lambda = c/f_c$ is the wavelength of the carrier, $f_c$ is the carrier frequency, and $c$ is the velocity of light in vacuum. The steering vector for the array is given by
\begin{equation}\label{equ:steeringVec}
	{\bf{a}}\left( {{{\bf{p}}_k}} \right) = [ {{a_{p,q}}\left( {{{\bf{p}}_k}} \right)} ]\left| {_{p = 0,1,\cdots,P - 1;q = 0,1,\cdots,Q - 1}}\right.,
\end{equation}
where $\mathbf{a}(\mathbf{p}_k) \in \mathbb{C}^{PQ \times 1}$. The sizes of the antenna arrays of the user and BS are ${P_r} \times {Q_r}$ and ${P_t} \times {Q_t}$, respectively.
 
\subsection{UL ISAC Channel Model} \label{subsec:JCAS_channel}
In this paper, we assume there exists a strong LoS path and several weak reflective multipaths in the mmWave channel between UE and BS. The range and AoA of UE contained in the channel are also the sensing parameters to be estimated for localization.
The UL ISAC channel response at the $n$th subcarrier of the $m$th packet can be expressed as~\cite{Zhang2022ISAC}
\begin{equation}\label{equ:H_c^U}
	{\bf{H}}_{n,m}{\rm{ = }}\sum\limits_{k = 0}^{K - 1} {\left[ \begin{array}{l}
			{b_{C,k}}{e^{j2\pi {{f_{d,k}}} mT_s^p}}{e^{ - j2\pi n\Delta {f} {{\tau _{k}}} }}\\
			\times {\bf{a}}( {{\bf{p}}_{R,k}^U} ){{\bf{a}}^T}( {{\bf{p}}_{T,k}^U} )
		\end{array} \right]},
\end{equation}
where ${\bf{H}}_{n,m} \in \mathbb{C}^{{P_t}{Q_t} \times {P_r}{Q_r}}$, $k = 0$ is for the channel response of the LoS path, and $k \in \{1, \cdots, K-1\}$ is for the $k$th non-line-of-sight (NLoS) path; ${\bf{p}}_{R,k}^U$ and ${\bf{p}}_{T,k}^U$ are the corresponding AoA and AoD, respectively; ${\bf{a}}( {{\bf{p}}_{R,k}^U} ) \in \mathbb{C}^{{P_t}{Q_t}\times 1}$ and ${\bf{a}}( {{\bf{p}}_{T,k}^U} ) \in \mathbb{C}^{{P_r}{Q_r}\times 1}$ are the steering vectors for UL reception and transmission, respectively; ${f_{d,0}} = \frac{{{v_{0,1}}}}{\lambda }$ and ${\tau _{c,0}} = \frac{{{r_{0,1}}}}{c}$ are the Doppler shift and time delay between UE and BS of the LoS path, respectively, with ${v_{0,1}}$ and ${r_{0,1}}$ being the corresponding radial relative velocity and the distance, respectively; ${f_{d,k}} = {f_{k,1}} + {f_{k,2}}$ and ${\tau _{k}} = {\tau _{k,1}} + {\tau _{k,2}}$ are the aggregate Doppler shift and time delay of the $k$th NLoS path, respectively; ${f_{k,1}} = \frac{{{v_{k,1}}}}{\lambda }$ and ${f_{k,2}} = \frac{{{v_{k,2}}}}{\lambda }$ are the Doppler frequencies between UE and the $k$th scatterer, and between the $k$th scatterer and the BS, respectively, with ${v_{k,1}}$ and ${v_{k,2}}$ being the corresponding radial velocities; ${\tau _{k,1}} = \frac{{{r_{k,1}}}}{c}$ and ${\tau _{k,2}} = \frac{{{r_{k,2}}}}{c}$ are the time delays between UE and the $k$th scatterer, and between BS and the $k$th scatterer, respectively, with ${r_{k,1}}$ and ${r_{k,2}}$ being the corresponding distances. Moreover, ${b_{C,0}} = \sqrt {\frac{{{\lambda ^2}}}{{{{(4\pi {r_{0,1}})}^2}}}}$ and ${b_{C,k}} = \sqrt {\frac{{{\lambda ^2}}}{{{{\left( {4\pi } \right)}^3}{r_{k,1}}^2{r_{k,2}}^2}}} {\beta _{C,k}}$ are the attenuations of the LoS and NLoS paths, respectively, where ${\beta _{C,k}}$ is the reflecting factor of the $k$th scatterer, following $\mathcal{CN}(0,\sigma _{C\beta ,k}^2)$~\cite{Rodger2014principles}. 

\subsection{CSI estimation}

The CSI can be initially estimated based on, e.g., the LS method using the training sequences in the preamble~\cite{2010MIMO}. We assume that a transmit beamforming (BF) with the BF vector ${\bf{w}}_{T}$ is used. Omitting the details of training sequences and estimation method, we present the UL CSI estimated by the LS method at the $n$th subcarrier of the $m$th packet as
\begin{equation}\label{equ:h_c_U_bar}
	\begin{array}{l}
		{{{\bf{\hat h}}}_{n,m}} = {{\bf{h}}_{n,m}} + {{{\bf{ n}}}_{n,m}} \in {\mathbb{C}^{{P_t}{Q_t} \times 1}}\\
		= \sqrt {P_t^U} {{\bf{H}}_{n,m}}{{\bf{w}}_{T}}{\zeta _{f,m}}{\zeta _{\tau,m}} + {{{\bf{ n}}}_{n,m}}\\
		= \sqrt {P_t^U} \sum\limits_{k = 0}^{K - 1} {\left[ \begin{array}{l}
				{b_{C,k}}{e^{j2\pi m{{T_s^{p}}}[{f_{d,k}} + {\delta _f}\left( m \right)]}}\\
				\times {e^{ - j2\pi n\Delta f[{\tau _{k}} + {\delta _\tau }\left( m \right)]}}\\
				\times {\chi _{T,k}}{\bf{a}}( {{\bf{p}}_{R,k}^U} )
			\end{array} \right]}  + {{{\bf{ n}}}_{n,m}},
	\end{array}
\end{equation}
where ${\bf{h}}_{n,m} = \sqrt {P_t^U} {\bf{H}}_{n,m}{\bf{w}}_{T}{\zeta _{f,m}}{\zeta _{\tau,m}}$ is the equivalent channel response, ${\zeta _{f,m}} = {e^{j2\pi m{T_s^{p}}{\delta _f}\left( m \right)}}$ and ${\zeta _{\tau,m}} = {e^{ - j2\pi n\Delta f{\delta _\tau }\left( m \right)}}$ are the phase shifts due to carrier frequency offset (CFO) and TO, denoted by ${\delta _f}\left( m \right)$ and ${\delta _\tau }\left( m \right)$, respectively; ${\delta _f}\left( m \right)$ and ${\delta _\tau }\left( m \right)$ are random time-varying parameters, following Gaussian distribution with zero mean and variances $\sigma_f^2$ and $\sigma _\tau^2$, respectively; $T_s^p$ is the time interval between two CSI estimates; ${\bf{ n}}_{n,m} \in \mathbb{C}^{{P_t}{Q_t} \times 1}$ is the combined noise that contains complex Gaussian noise, and each element of
${\bf{n}}_{n,m}$ follows $\mathcal{CN}(0,\sigma _N^2)$; ${\bf{w}}_{T}$ is the transmit BF vector with ${\left\| {{{\bf{w}}_{T}}}  \right\|_2} = 1$, and $\chi _{T,k} = {{\bf{a}}^T}( {{\bf{p}}_{T,k}^U} ) {{\bf{w}}_{T}}$ is the transmit BF gain. 

\subsection{Received Data Signals}
The $i$th received OFDM symbol at the $n$th subcarrier of the $m$th packet can be expressed as
\begin{equation}\label{equ:y_nmi}
	{\bf{y}}_{n,m}^i = {{\bf{h}}_{n,m}}d_{n,m}^i + {\bf{n}}_{n,m}^i \in {\mathbb{C}^{{P_t}{Q_t} \times 1}},
\end{equation}
where $d_{n,m}^i$ is the $i$th transmitted OFDM data symbol at the $n$th subcarrier of the $m$th packet, ${\bf{n}}_{n,m}^i$ is the Gaussian noise vector with each element following $\mathcal{CN}(0,\sigma _N^2)$, and ${{\bf{h}}_{n,m}}$ is the equivalent channel response in \eqref{equ:h_c_U_bar}.

\section{Joint Single-base Localization and CSI Enhancement in Preamble Period}\label{sec:JCAS_sensing}

We present the UL single-base joint localization and communication enhancement scheme as shown in Fig.~\ref{fig: UL communication}. This section introduces the signal processing procedures using the preamble signals. BS conducts initial preamble-based CSI (P-CSI) estimation using, e.g., the LS channel estimation method. Based on the singular value decomposition (SVD) of the estimated P-CSIs, we can estimate AoAs without being impacted by TO and CFO using a refined 2D MUSIC-based AoA estimation method and improve the accuracy of CSI estimation in the process of AoA estimation. Next, using the AoA estimates, we construct a spatial filter to separate CSIs with different AoAs. We then introduce the MUSIC-based range estimation method to estimate the ranges exploiting the separated CSIs corresponding to the estimated AoAs. We prove that the MUSIC-based range estimation method can suppress the noise-like TO terms in the CSI to enhance the accuracy of range estimation. Finally, we can localize the UE with the spherical coordinates determined by the estimated AoA and range.


\begin{algorithm}[!t]
	\caption{2D two-step Newton descent minimum searching method~\cite{2023XuJCAS}}
	\label{alg:Two-step_descent}
	\KwIn{The range of $\varphi $: ${\Phi _\varphi }$; the range of $\theta $: ${\Phi _\theta }$; the number of grid points: ${N_i}$; the maximum iteration round $ind_{max}$; the MUSIC spectrum function: $f(\bf p)$.
	}
	\KwOut{Estimation results: ${\bf{\Theta }}\!\! =\!\! { {\{ {{{{\bf{\hat p}}}_k}} \}} |_{k \in \{ {0,...,{N_s} - 1} \}}}$.}
	\textbf{Initialize: } \\
	1) ${\Phi _\varphi }$ and ${\Phi _\theta }$ are both divided evenly into ${N_i} - 1$ pieces with ${N_i}$ grid points to generate grid ${\hat \Phi _\varphi }$ and ${\hat \Phi _\theta }$.\\
	2) Set a null space ${\bf{\Theta }}$.\\
	\textbf{Process: } \\
	\textbf{Step 1}: \ForEach{${{\bf{p}}_{i,j}} \in {\hat \Phi _\varphi } \times {\Phi _\theta }$}
	{
		Calculate the spatial spectrum as ${\bf{S}}$, where ${[{\bf{S}}]_{i,j}} = [f\left( {{{\bf{p}}_{i,j}}} \right)]^{-1}$.
	}
	\textbf{Step 2}: Search the maximal values of ${\bf{S}}$ to form the set ${\bar \Theta _d}$.
	
	\textbf{Step 3}: Derive the Hessian matrix and the gradient vector of $f( {\bf{p}} )$ as ${\bf{H}}_{\bf{p}}\left( {\bf{p}} \right)$ and ${\nabla _{\bf{p}}}f\left( {\bf{p}} \right)$, respectively. 
	
	\textbf{Step 4}: \ForEach{${{\bf{p}}_{i,j}} \in {\bar \Theta _d}$}{
		\textit{k}=1\;
		${{\bf{p}}^{( 0 )}} = {{\bf{p}}_{i,j}}$\;
		${{\bf{p}}^{( k )}} = {{\bf{p}}^{\left( {k - 1} \right)}} - {[ {{{\bf{H}}_{\bf{p}}}( {{{\bf{p}}^{( {k - 1} )}}} )} ]^{ - 1}}{\nabla _{\bf{p}}}f( {{{\bf{p}}^{( {k - 1} )}}} )$\;
		\While{$\| {{{\bf{p}}^{\left( k \right)}} - {{\bf{p}}^{( {k - 1} )}}} \| > \varepsilon $ and $k \le ind_{max}$}
		{
			${{\bf{p}}^{( k )}}= {{\bf{p}}^{( {k - 1} )}} - {[ {{{\bf{H}}_{\bf{p}}}( {{{\bf{p}}^{( {k - 1} )}}} )} ]^{ - 1}}{\nabla _{\bf{p}}}f( {{{\bf{p}}^{( {k - 1} )}}} )$\;
			
		}
		${{\bf{p}}^{\left( k \right)}}$ is put into output set ${\bf{\Theta }}$\;
	}
\end{algorithm}

\subsection{Joint AoA Estimation and CSI Enhancement} \label{sec:AoA_estimation_CSI}

We choose the MUSIC-based method to estimate AoAs as it can work with non-equally spaced measurements in time, frequency, and spatial domains, offering great flexibility to resource allocation. Such non-regular measurements are common in communication systems.

\subsubsection{AoA Estimation}
Stacking all $M_s \times N_c$ CSI estimates (from $N_c$ subcarriers and $M_s$ packets), we obtain the CSI matrix as
\begin{equation}\label{equ:H_est}
	{{\bf{\hat H}}_C} = {\bf{H}} + {\bf{\hat N}} \in {\mathbb{C}^{{P_t}{Q_t} \times N_c M_s}},
\end{equation} 
where ${\bf{H}}$ is the actual value corresponding to ${\bf{\hat H}}_C$, and the $[(m-1)N_c + n]$th columns of ${\bf{\hat H}}_C$, ${\bf{H}}$, and ${\bf{\hat N}}$ are ${{{\bf{\hat h}}}_{n,m}}$, ${{\bf{h}}_{n,m}}$, and ${{{\bf{ n}}}_{n,m}}$, respectively.

Construct the correlation matrix of ${{\bf{\hat H}}_C}$ as
\begin{equation}\label{equ:R_x}
{\bf{R}}_{\bf{x}}{\rm{ = }}{{[ {{\bf{\hat H}}_C{{( {{\bf{\hat H}}_C} )}^H}} ]} \mathord{/
{\vphantom {{[ {{\bf{\hat H}}_C{{( {{\bf{\hat H}}_C} )}^H}} ]} {(M_s N_c)}}} 
\kern-\nulldelimiterspace} {(M_s N_c)}}.
\end{equation} 

Note that ${{\bf{\hat H}}_C}{({{\bf{\hat H}}_C})^H} = \sum\limits_{n,m}^{{N_c},{M_s}} {{{{\bf{\hat h}}}_{n,m}}{{({{{\bf{\hat h}}}_{n,m}})}^H}}$. According to the proof in \cite{Chen2023KF}, the AoA estimation is not prominently affected by the CFO and TO, which will also be shown in Section~\ref{sec:Simulation}. Next, we use a refined 2D MUSIC method~\cite{2023XuJCAS} to estimate the 2D AoA based on the SVD of ${\bf{\hat H}}_C$. The SVD of ${\bf{\hat H}}_C$ can be expressed as
\begin{equation}\label{equ:eigenvalue deomposition of Rx}
	{{\bf{\hat H}}_C} = {\bf{U\Sigma }}{{\bf{V}}^H} = \left[ {{{\bf{U}}_s},{{\bf{U}}_0}} \right]\left[ {\begin{array}{*{20}{c}}
			{{{\bf{\Sigma }}_s}}&0\\
			0&{{{\bf{\Sigma }}_0}}
	\end{array}} \right]\left[ {\begin{array}{*{20}{c}}
			{{{\bf{V}}_s}^H}\\
			{{{\bf{V}}_0}^H}
	\end{array}} \right],
\end{equation}
where ${{\bf{U}}_s}$ and ${{\bf{U}}_0}$ are the block matrices in the left singular matrix that are correlated to the signal and noise subspace, respectively; ${{\bf{V}}_s}$ and ${{\bf{V}}_0}$ are the block matrices in the right singular matrix that are correlated to the signal and noise subspace, respectively; and ${{\bf{\Sigma }}_s}$ and ${{\bf{\Sigma }}_0}$ are the diagonal matrices that are composed of the signal and noise singular values, respectively. The number of the identified AoAs is the rank of ${{\bf{\Sigma }}_s}$ and can be ascertained by estimating the rank of the signal space from the differential vector derived from the singular values (or eigenvalues) of the signal matrix, which is presented in the appendix in~\cite{2023XuJCAS}. 

We use $N_A$ to denote the number of estimated AoAs. Then, the noise subspace of ${{\bf{\hat H}}_C}$ is ${\bf{U}}_0 = {\left[ {{\bf{U}}} \right]_{:,N_A + 1:{P_t}{Q_t}}}$, and we can formulate the angle spectrum function as
\begin{equation}\label{equ:angle_spectrum_function}
	f_a( {\bf{p}} ) = {{\bf{a}}^H}( {\bf{p}} ){\bf{U}}_0{( {{\bf{U}}_0} )^H}{\bf{a}}( {\bf{p}} ),
\end{equation}
where ${\bf{a}}\left( {\bf{p}} \right)$ is given in~\eqref{equ:steeringVec}. 
The minimum points of $f_a( {\bf{p}} )$ correspond to the AoAs to be estimated. We exploit a 2D two-step Newton descent method~\cite{2023XuJCAS} to obtain the minimum points of $f_a( {\bf{p}} )$, which is shown in \textbf{Algorithm~\ref{alg:Two-step_descent}}. To identify the minimum of $f_a( {\bf{p}} )$, we substitute $f( {\bf{p}} )$, ${\bf{H}}_{\bf{p}}\left( {\bf{p}} \right)$, and ${\nabla _{\bf{p}}}f\left( {\bf{p}} \right)$ in \textbf{Algorithm~\ref{alg:Two-step_descent}} with \eqref{equ:angle_spectrum_function}, Hessian matrix and the gradient vector of $f_a( {\bf{p}} )$, respectively.
Since the gain of LoS path is significantly larger than those of the weak NLoS paths, the estimated angle with the largest $f_a^{ - 1}({\bf{p}})$ should be the AoA of UE, which is denoted by ${\bf{\hat p}}_{R,0}^U = ( {{{\hat \varphi }_0},{{\hat \theta }_0}} )$.

Based on the eigenvalue vector, denoted by ${{\bf{v}}_s} = {\rm{vec}}\left( {{{\bf{\Sigma }}_x}} \right)$, obtained in the MUSIC process, we can also estimate the variance of ${{{\bf{\bar n}}}_{n,m}}$ as $\hat \sigma _N^2$. According to \cite{HAARDT2014651}, ${{\bf{v}}_s} \in {\mathbb{C}^{{P_t}{Q_t} \times 1}}$ can be expressed as
\begin{equation}\label{equ:vs}
	{[{{\bf{v}}_s}]_i} = \left\{ {\begin{array}{*{20}{l}}
			{{P_i} + \sigma _N^2, for \; i \le {N_A}},\\
			{\sigma _N^2, for \; i > {N_A}},
	\end{array}} \right.
\end{equation}
where ${P_i}$ is the power of the $i$th incident signal. Therefore, we obtain the estimation of the noise power as
\begin{equation}\label{equ:noise_est}
	\hat \sigma _N^2 = {{\sum\limits_{i = N_A + 1}^{{P_t}{Q_t}} {{{[ {{{\bf{v}}_s}} ]}_i}} } \mathord{/
			{\vphantom {{\sum\limits_{i = N_A + 1}^{{P_t}{Q_t}} {{{\left[ {{{\bf{v}}_s}} \right]}_i}} } {( {{P_t}{Q_t} - N_A} )}}} 
			\kern-\nulldelimiterspace} {( {{P_t}{Q_t} - N_A} )}}.
\end{equation}

\subsubsection{CSI Enhancement} 
Since the effective rank of ${\bf{\hat H}}_C$ is estimated as $N_A$, we obtain ${{\bf{\Sigma }}_s} = {\left[ {\bf{\Sigma }} \right]_{1:{N_A},1:{N_A}}}$, ${{\bf{U}}_s} = {\left[ {\bf{U}} \right]_{1:{N_A},:}}$, and ${{\bf{V}}_s} = {\left[ {\bf{V}} \right]_{1:{N_A},:}}$. The enhanced CSI estimation that eliminates the noise subspace can be expressed as
\begin{equation}\label{equ:H_bar_1enhancement}
	{\bf{\bar H}} = {{\bf{U}}_s}{{\bf{\Sigma }}_s}{{\bf{V}}_s}^H = {\bf{H}} + {\bf{\bar N}},
\end{equation}
where ${\bf{\bar N}}$ is the colored noise blended into the signal subspace. Here, ${\bf{\bar N}}$ has the following features. First, ${\bf{\bar N}}$ is irrelevant to ${\bf{H}}$, i.e., $E\left\{ {{\bf{H}}{{{\bf{\bar N}}}^H}} \right\} = {\bf{0}}$. Second, the eigenvalues of the autocorrelation of ${\bf{\bar N}}$ are highly centralized, and there is $E\left\{ {{\bf{\bar N}}{{{\bf{\bar N}}}^H}} \right\} \approx {{\bf{U}}_s} \sigma _N^2{{\bf{U}}_s}^H$~\cite{Moor1993}. Based on the above two features, we can obtain the autocorrelation of ${{\bf{\bar H}}}$ and $\bf{H}$, respectively, as 
\begin{equation}\label{equ:EHH_bar}
	\begin{array}{l}
		{{\bf{R}}_{{\bf{\bar H}}}} = E\left\{ {{\bf{\bar H}}{{{\bf{\bar H}}}^H}} \right\}
		= {{\bf{U}}_s}{{\bf{\Sigma }}_s}^2{{\bf{U}}_s}^H
	\end{array},
\end{equation}
and
\begin{equation}\label{equ:EHH}
	\begin{array}{l}
		{{\bf{R}}_{\bf{H}}} = E\left\{ {{\bf{H}}{{\bf{H}}^H}} \right\}\\
		= {{\bf{R}}_{{\bf{\bar H}}}} - E\left\{ {{\bf{\bar N}}{{{\bf{\bar N}}}^H}} \right\}\\
		= {{\bf{U}}_s}{{\bf{\Sigma }}_s}^2{{\bf{U}}_s}^H - {{\bf{U}}_s}\sigma _N^2{{\bf{U}}_s}^H\\
		= {{\bf{U}}_s}\left( {{{\bf{\Sigma }}_s}^2 - \sigma _N^2{\bf{I}}} \right){{\bf{U}}_s}^H
	\end{array}.
\end{equation}

Next, we form the following optimization problem to find the optimal combination of ${\bf{\bar H}}$, denoted by ${{\bf{\bar H}}_u} = {\bf{B\bar H}}$, to suppress the colored noise in ${\bf{\bar H}}$.
\begin{equation}\label{equ:optimize_H}
	\begin{array}{l}
		{\bf{B}} = \arg \mathop {\min }\limits_{\bf{B}} E \left\| {{\bf{B\bar H}} - {\bf{H}}} \right\|_F^2\\
		s.t. \; {\rm{ }}E\left\{ {{\bf{H}}{{\bf{H}}^H}} \right\} = {{\bf{U}}_s}\left( {{{\bf{\Sigma }}_s}^2 - \sigma _N^2{\bf{I}}} \right){{\bf{U}}_s}^H
	\end{array}.
\end{equation}

The detailed derivation to the \eqref{equ:optimize_H} is provided in \textbf{Appendix~\ref{appendix:optimize_H}}, and the optimal solution is 
\begin{equation}\label{equ:optimal_B}
	{{\bf{B}}_{opt}} = {{\bf{R}}_{\bf{H}}}{\left( {{{\bf{R}}_{{\bf{\bar H}}}}} \right)^{ - 1}}.
\end{equation}

Combining \eqref{equ:H_bar_1enhancement}, \eqref{equ:EHH_bar}, \eqref{equ:EHH}, and \eqref{equ:optimal_B}, we obtain the enhanced CSI estimation as
\begin{equation}\label{equ:H_u_bar}
	\begin{array}{l}
		{{{\bf{\bar H}}}_u} = {{\bf{R}}_{\bf{H}}}{\left( {{{\bf{R}}_{{\bf{\bar H}}}}} \right)^{ - 1}}{\bf{\bar H}}\\
		= {{\bf{U}}_s}\left( {{{\bf{\Sigma }}_s}^2 - \sigma _N^2{\bf{I}}} \right){{\bf{\Sigma }}_s}^{ - 2}{{\bf{U}}_s}^H{\bf{\bar H}}\\
		= {{\bf{U}}_s}\left( {{{\bf{\Sigma }}_s}^2 - \sigma _N^2{\bf{I}}} \right){{\bf{\Sigma }}_s}^{ - 2}{{\bf{U}}_s}^H{{\bf{U}}_s}{{\bf{\Sigma }}_s}{{\bf{V}}_s}^H\\
		= {{\bf{U}}_s}\left( {{{\bf{\Sigma }}_s}^2 - \sigma _N^2{\bf{I}}} \right){{\bf{\Sigma }}_s}^{ - 1}{{\bf{V}}_s}^H
	\end{array},
\end{equation}
where $\sigma _N^2$ can be replaced by $\hat \sigma _N^2$ to obtain the approximate value of ${{{\bf{\bar H}}}_u}$. Moreover, ${{\bf{\bar H}}_u}$ can be further expressed as
${{\bf{\bar H}}_u} = {\bf{H}} + {{\bf{\bar N}}_u}$, where ${{\bf{\bar N}}_u}$ is the suppressed noise.

Next, ${{{\bf{\bar H}}}_u}$ can be utilized for enhancing communication demodulation.

\subsubsection{Spatial Filtering} \label{sec:Spatial_Filter}
In order to further estimate the range of UE for localization, we apply a baseband spatial filter for receiving the signals in ${\bf{\hat p}}_{R,0}^U = ( {{{\hat \varphi }_0},{{\hat \theta }_0}} )$. The BF vector for receiving the signals from the AoA of ${\bf{\hat p}}_{R,0}^U$ is generated with the LS method as~\cite{Zhang2019JCRS}
\begin{equation}\label{equ:w_RL}
	{{\bf{w}}_{R}} = \frac{{{{\left[ {{{\bf{a}}^T}\left( {{\bf{\hat p}}_{R,0}^U} \right)} \right]}^\dag }}}{{\sqrt {\left\| {{{\left[ {{{\bf{a}}^T}\left( {{\bf{\hat p}}_{R,0}^U} \right)} \right]}^\dag }} \right\|_2^2} }} \in \mathbb{C}^{P_t Q_t \times 1},
\end{equation}
where $[\cdot]^{\dag}$ represents the pseudo-inverse of a matrix.

Using ${{\bf{w}}_{R}}$ to filter ${{{\bf{\bar H}}}_u}$, we obtain the filtered vector ${{\bf{\bar h}}_R} = {\left( {{{\bf{w}}_R}} \right)^H} {{\bf{\bar H}}_u} \in \mathbb{C}^{1 \times N_c M_s}$, and the $[(m-1)N_c + n]$th column of ${{\bf{\bar h}}_R}$ can be expressed as 
\begin{equation}\label{equ:h_nm_l}
	\begin{array}{*{20}{l}}
		{{{\bar h}_{n,m}}{\rm{ = }}{{\left( {{{\bf{w}}_R}} \right)}^H}{{\bf{h}}_{n,m}} + {{\bar n}_{n,m}}}\\
		{ = \sqrt {P_t^U} {b_{C,0}}{\chi _{T,0}}{\chi _{R,0}}{e^{j2\pi mT_s^p{{\tilde f}_{d,0,m}}}}{e^{ - j2\pi n\Delta f{{\tilde \tau }_{0,m}}}} + {{\bar n}_{n,m}}},
	\end{array}
\end{equation} 
where ${{\tilde f}_{d,0,m}} = {f_{d,0}} + {\delta _f}\left( m \right)$, ${{\tilde \tau }_{0,m}} = {\tau _{0}} + {\delta _\tau }\left( m \right)$,  and ${\chi _{R,0}} = {( {{{\bf{w}}_{R}}} )^H}{\bf{a}}( {{\bf{p}}_{R,0}^U} )$ is the directional receiving gain of the signals. Moreover, the equivalent interference-plus-noise term is
\begin{equation}\label{equ:n_nm_l}
	\begin{array}{*{20}{l}}
		{\bar n_{n,m}}\\
		{ = {{\left( {{{\bf{w}}_R}} \right)}^H} {{{\bf{\bar n}}}_{n,m}} + \sum\limits_{k = 1}^{K - 1}  \left[ {\begin{array}{*{20}{l}}
					{{e^{j2\pi mT_s^p{{\tilde f}_{d,k,m}}}}{e^{ - j2\pi n\Delta f{{\tilde \tau }_{k,m}}}}}\\
					{ \times \sqrt {P_t^U} {b_{C,k}}{\chi _{T,k}}{\chi _{R,I,k}}}
			\end{array}} \right]},
	\end{array}
\end{equation} 
where ${\chi _{R,I,k}} = {\left( {{{\bf{w}}_{R}}} \right)^H}{\bf{a}}( {{\bf{p}}_{R,k}^U} ) (k \ne 0)$ is the receive gain of interference from the $k$th weak NLoS path, and ${{{\bf{\bar n}}}_{n,m}} = {\left[ {{{{\bf{\bar N}}}_u}} \right]_{:,(m - 1){N_c}{\rm{ + }}n}}$. It is easy to see that $\left\| {{\chi _{R,I,k}}} \right\|_2^2 \ll \left\| {{\chi _{R,0}}} \right\|_2^2$.

\subsection{MUSIC-based Range Estimation} \label{MUSIC_range_estimation}

Reshape ${{\bf{\bar h}}_R}$ to form a matrix ${{\bf{\bar H}}_R} \in {\mathbb{C}^{{N_c} \times {M_s}}}$ for range estimation, where ${\left[ {{{{\bf{\bar H}}}_R}} \right]_{n,m}} = {\bar h_{n,m}}$. According to \eqref{equ:h_nm_l}, ${{\bf{\bar H}}_R}$ has steering vector-like expressions, i.e., ${e^{ - j2\pi n\Delta f{{\tilde \tau }_{k,m}}}}$ and ${e^{j2\pi m{T_s^p}{{\tilde f}_{d,k,m}}}}$, we can construct the range and Doppler steering vectors, respectively, as
\begin{equation}\label{equ:a_r}
	{{\bf{a}}_{{\bf{r}},m}} = [{e^{ - j2\pi n\Delta f\frac{{{{\tilde r}_m}}}{c}}}]{|_{n = 0,1,...,{N_c} - 1}} \in {\mathbb{C}^{{N_c} \times 1}},
\end{equation}
\begin{equation}\label{equ:a_f}
	{{\bf{a}}_{{\bf{f}}}} = [{e^{j2\pi mT_s^p{{\tilde f}_{d,0,m}}}}]{|_{m = 0,1,...,{M_s} - 1}} \in {\mathbb{C}^{{M_s} \times 1}},
\end{equation}
where ${\tilde r_{m}} = {\tilde \tau _{0,m}} \times c$ is the range of UE plus the offset due to TO. 

Then, based on \eqref{equ:h_nm_l}, the $m$th column of ${{\bf{\bar H}}_R}$ can be expressed by \eqref{equ:a_r} and \eqref{equ:a_f} as
\begin{equation}\label{equ:H_R_col}
	{\left[ {{{{\bf{\bar H}}}_R}} \right]_{:,m}} = {S_0}{{\bf{a}}_{{\bf{r}},m}}{\left[ {{{\bf{a}}_{\bf{f}}}} \right]_m} + {\left[ {{{{\bf{\bar N}}}_R}} \right]_{_{:,m}}},
\end{equation}
where ${S_0} = \sqrt {P_t^U} {b_{C,0}} {\chi _{R,0}}{\chi _{T,0}}$, and ${\left[ {{{{\bf{\bar N}}}_R}} \right]_{n,m}} = {{\bar n}_{n,m}}$. We can use the MUSIC-based range estimation method \cite{2023XuJCAS} to estimate the range of UE based on the autocorrelation of ${{\bf{\bar H}}_R}$, as given by 
\begin{equation}\label{equ:R_r}
	\begin{array}{l}
		{{\bf{R}}_r} = \frac{1}{{{M_s}}}{{{\bf{\bar H}}}_R}{\left( {{{{\bf{\bar H}}}_R}} \right)^H}\\
		= \frac{1}{{{M_s}}}\sum\limits_{m = 0}^{{M_s} - 1} {{{\left[ {{{{\bf{\bar H}}}_R}} \right]}_{:,m}}{{\left( {{{\left[ {{{{\bf{\bar H}}}_R}} \right]}_{:,m}}} \right)}^H}}.
	\end{array}
\end{equation}
Here, we further prove an important feature of the MUSIC-based range estimation method which is beneficial to localization under clock asynchronism in \textbf{Proposition~\ref{proposition_MUSIC_range}}.
\begin{proposition} \label{proposition_MUSIC_range}
	The MUSIC-based range estimation method can suppress the noise-like phase-shift terms due to TO.
	\begin{proof}
		The detailed proof is provided in \textbf{Appendix~\ref{appendix:MUSIC}}. 
	\end{proof}
\end{proposition}

Next, we briefly introduce the steps of MUSIC-based range estimation method. By applying eigenvalue decomposition (EVD) to ${{\bf{R}}_r}$, we obtain
\begin{equation}\label{equ:eig}
	{{\bf{R}}_r} = {{\bf{U}}_r}{{\bf{\Sigma }}_r}{\left( {{{\bf{U}}_r}} \right)^H},
\end{equation}
where ${{\bf{\Sigma }}_r}$ is the eigenvalue diagonal matrix, and ${{\bf{U}}_r}$ is the corresponding eigen matrix. The noise subspace of ${{\bf{R}}_r}$ for estimating the range of UE is ${{\bf{U}}_{rN}} = {\left[ {{{\bf{U}}_r}} \right]_{:,2:{N_c}}}$. Based on ${{\bf{U}}_{rN}}$, we can obtain the range spectrum function as
\begin{equation}\label{equ:f_rr}
	{f_r}(r) = {{\bf{a}}_r}{(r)^H}{{\bf{U}}_{rN}}{({{\bf{U}}_{rN}})^H}{{\bf{a}}_r}(r),
\end{equation}
where ${{\bf{a}}_r}\left( r \right) = [{e^{ - j2\pi n\Delta f\frac{r}{c}}}]{|_{n = 0,1,...,{N_c} - 1}} \in {\mathbb{C}^{{N_c} \times 1}}$ is the range steering vector.

The minimal point of $f_r\left( r \right)$ corresponds to the range to be estimated, and \textbf{Algorithm~\ref{alg:Two-step_descent}} can be used to identify the minimal point of $f_r\left( r \right)$. Note that $f\left( {{{\bf{p}}}} \right)$, ${\nabla _{\bf{p}}}f\left( {{{\bf{p}}}} \right)$ and ${{{\bf{H}}_{\bf{p}}}\left( {{{\bf{p}}}} \right)}$ in \textbf{Algorithm~\ref{alg:Two-step_descent}} are replaced by \eqref{equ:f_rr}, $\frac{{\partial {f_r}\left( r \right)}}{{\partial r}}$, and  $\frac{{{\partial ^2}{f_r}\left( r \right)}}{{{\partial ^2}r}}$ for range estimation, respectively. The estimated range of UE is denoted by $\hat r_0$.

\subsection{Localization of UE}

Since $\hat r_0$ corresponds to the estimated AoA of UE, ${\bf{\hat p}}_{R,0}^U = ( {{{\hat \varphi }_0},{{\hat \theta }_0}} )$, as shown in \eqref{equ:w_RL} and \eqref{equ:h_nm_l}, $\hat r_0$ and ${\bf{\hat p}}_{R,0}^U$ form the polar cordinate of UE. Therefore, the Cartesian coordinate of UE can be expressed as
\begin{equation}\label{equ:loc_UE}
	{{\bf{\Omega }}_0} = ({\hat r_0}\sin {\hat \theta _0}\cos {\hat \varphi _0},{\hat r_0}\sin {\hat \theta _0}\sin {\hat \varphi _0},{\hat r_0}\cos {\hat \theta _0}).
\end{equation}

\section{Joint CSI and Data Signals-based Localization Scheme} \label{sec:Joint_CSI_data_localization}
In this section, we first demodulate the communication data signals based on the enhanced CSI obtained in Section \ref{sec:AoA_estimation_CSI}. Then, we propose a joint CSI and data signals-based localization scheme which exceeds the sensing accuracy of the localization scheme based on merely CSI as presented in Section \ref{sec:JCAS_sensing}.

\subsection{Data Signal Demodulation}~\label{sec:DSD_E_DSI}

The $[(m - 1){N_c} + n]$th column of ${{\bf{\bar H}}_u}$ is the enhanced CSI estimates of ${{\bf{h}}_{n,m}}$. We use ${{\bf{\bar h}}_{n,m}} = {\left[ {{{{\bf{\bar H}}}_u}} \right]_{:,(m - 1){N_c} + n}}$ to demodulate the received data signals and estimate the transmit data symbols. By applying the low-complexity zero-forcing to conduct channel equalization to ${\bf{y}}_{n,m}^i$, we obtain
\begin{equation}\label{equ:y_nmi_CE}
	y_{n,m}^i = {\left( {{{{\bf{\bar h}}}_{n,m}}} \right)^\dag }{\bf{y}}_{n,m}^i.
\end{equation}

Then, decoding the transmit data symbols based on $y_{n,m}^i$ using the maximum-likelihood (ML) criterion, the estimation of the transmit data symbol at the $n$th subcarrier of the $i$th OFDM symbol of the $m$th packet can be expressed as 
\begin{equation}\label{equ:d_nmi_CE}
	\hat d_{n,m}^i = \mathop {\arg \min }\limits_{d \in {\Theta _{\rm QAM}}} {\left| {y_{n,m}^i - d} \right|^2},
\end{equation}
where ${{\Theta _{\rm QAM}}}$ is the used quadrature amplitude modulation (QAM) constellation.

Next, using the decoded data symbols and the received data signals, we can obtain the data-based CSI (D-CSI) as 
\begin{equation}\label{equ:h_nm_i}
	\begin{array}{l}
		{\bf{\hat h}}_{n,m}^i = \frac{{{\bf{y}}_{n,m}^i}}{{\hat d_{n,m}^i}}\\
		= {{\bf{h}}_{n,m}}\frac{{d_{n,m}^i}}{{\hat d_{n,m}^i}} + {\bf{\hat n}}_{n,m}^i\\
		= {{\bf{h}}_{n,m}}\frac{1}{{1 + \psi _{n,m}^i}} + {\bf{\hat n}}_{n,m}^i,
	\end{array}
\end{equation}
where $\psi _{n,m}^i = \frac{{e_{n,m}^i}}{{d_{n,m}^i}}$, $e_{n,m}^i = \hat d_{n,m}^i - d_{n,m}^i$ is the symbol decoding error, and ${\bf{\hat n}}_{n,m}^i = \frac{{{\bf{n}}_{n,m}^i}}{{\hat d_{n,m}^i}}$ is the transformed Gaussian noise. We can see that when the transmit data symbol is decoded without error, i.e., $\psi _{n,m}^i = 1$, the estimated D-CSI is equal to the CSI estimated by LS method. 

Stack all the estimated D-CSI ${\bf{\hat h}}_{n,m}^i$ at $N_c$ subcarriers of $M_s$ symbols to form a matrix ${{\bf{\hat H}}^d} \in {\mathbb{C}^{{P_t}{Q_t} \times  {P_s}{M_s}{N_c}}}$, where ${\left[ {{{{\bf{\hat H}}}^d}} \right]_{:,[(i - 1){M_s}{N_c} + (m - 1){N_c} + n]}} = {\bf{\hat h}}_{n,m}^i$ is the ${[(i - 1){M_s}{N_c} + (m - 1){N_c} + n]}$th column of ${{{{\bf{\hat H}}}^d}}$, i.e., the D-CSI estimation at the $n$th subcarrier of the $i$th OFDM symbol of the $m$th packet. 

\subsection{Joint CSI and Data Signals-based Localization} \label{sec:JCDS_B_LOC}

In this subsection, we first present the joint CSI and data signals-based AoA and range estimation. Then, we introduce the single-base localization of UE based on the estimated AoA and range of UE. 

\subsubsection{Joint CSI and Data Signals-based AoA Estimation}
\label{sec:JCDS_AoA}
By stacking all the D-CSI and enhanced P-CSI into one matrix, we obtain
\begin{equation}\label{equ:H_tilde}
	{\bf{\tilde H}} = \left[ {{{{\bf{\bar H}}}_u},{{{\bf{\hat H}}}^d}} \right] \in {\mathbb{C}^{{P_t}{Q_t} \times ({P_s} + 1){M_s}{N_c}}},
\end{equation}
where the first $M_s N_c$ columns of ${\bf{\tilde H}}$ is ${{{\bf{\bar H}}}_u}$. According to~\eqref{equ:R_r}, we can see that the MUSIC-based method does not require the input signals to be consecutive in subcarriers and OFDM symbols. Therefore, the MUSIC-based method can still be used to estimate the AoA and range of UE based on the autocorrelation of ${\bf{\tilde H}}$. 

The autocorrelation of ${\bf{\tilde H}}$ can be expressed as
\begin{equation}\label{equ:R_H_tilde}
	\begin{array}{l}
		{{\bf{R}}_{{\bf{\tilde H}}}} = \frac{1}{{({P_s} + 1){M_s}{N_c}}}{\bf{\tilde H}}{{{\bf{\tilde H}}}^H}\\
		= \frac{1}{{({P_s} + 1){M_s}{N_c}}}[ {{{{\bf{\bar H}}}_u}{{( {{{{\bf{\bar H}}}_u}} )}^H} + {{{\bf{\hat H}}}^d}{{( {{{{\bf{\hat H}}}^d}} )}^H}} ]\\
		= \frac{1}{{({P_s} + 1){M_s}{N_c}}}[ { {{{\bf{\bar H}}}_u} {{( {{{{\bf{\bar H}}}_u}} )}^H} + \sum\limits_{n,m,i}^{} {{\bf{\hat h}}_{n,m}^i{{( {{\bf{\hat h}}_{n,m}^i} )}^H}} } ].
	\end{array}
\end{equation}

Then, we prove an important feature of ${{\bf{R}}_{{\bf{\tilde H}}}}$ that can improve the AoA sensing performance in \textbf{Proposition~\ref{proposition_AoA}}.

\begin{proposition} \label{proposition_AoA}
	${{\bf{R}}_{{\bf{\tilde H}}}}$ can be used in MUSIC-based method to enhance the coherent energy for AoA estimation.
	\begin{proof}
		The detailed proof is provided in \textbf{Appendix~\ref{appendix:AoA_enhancement}}. 
	\end{proof}
\end{proposition}
Subsequently, we present the MUSIC-based AoA estimation based on the joint CSI and data signals. 
By applying EVD to ${{\bf{R}}_{{\bf{\tilde H}}}}$, we obtain 
\begin{equation}\label{equ:R_H_EVD}
	{{\bf{R}}_{{\bf{\tilde H}}}} = {\bf{\tilde U\tilde \Sigma }}{{\bf{\tilde U}}^H}.
\end{equation}

Then, the noise subspace obtained from ${{\bf{R}}_{{\bf{\tilde H}}}}$ is ${{\bf{\tilde U}}_0} = {[ {{\bf{\tilde U}}} ]_{:,2:{N_A}}}$, and the angle spectrum function can be expressed as
\begin{equation}\label{equ:f_a_p}
	{\tilde f_a}({\bf{p}}) = {{\bf{a}}^H}({\bf{p}}){{\bf{\tilde U}}_0}{({{\bf{\tilde U}}_0})^H}{\bf{a}}({\bf{p}}).
\end{equation}

Subsequently, we use \textbf{Algorithm~\ref{alg:Two-step_descent}} to obtain the minimum points of ${\tilde f_a}({\bf{p}})$ to estimate the AoA of UE. To identify the minimum of ${\tilde f_a}({\bf{p}})$, we substitute $f( {\bf{p}} )$, ${\bf{H}}_{\bf{p}}\left( {\bf{p}} \right)$, and ${\nabla _{\bf{p}}}f\left( {\bf{p}} \right)$ in \textbf{Algorithm~\ref{alg:Two-step_descent}} with \eqref{equ:f_a_p}, Hessian matrix, and the gradient vector of ${\tilde f_a}({\bf{p}})$, respectively. Similar to the AoA estimation with pure CSI, the estimated angle with the largest ${\tilde f_a}^{ - 1}({\bf{p}})$ is the AoA estimate of UE, which is denoted by ${\bf{\tilde p}}_{R,0}^U = ( {{{\tilde \varphi }_0},{{\tilde \theta }_0}} )$.

\subsubsection{Joint CSI and Data Signals-based Range Estimation}

Applying a baseband spatial filter to aggregate the signals from the direction of ${\bf{\tilde p}}_{R,0}^U$ for estimating the range of UE, we obtain
\begin{equation}\label{equ:h_R}
	{{\bf{\tilde h}}_R} = {\left( {{{{\bf{\tilde w}}}_R}} \right)^H}{\bf{\tilde H}} \in {\mathbb{C}^{1 \times ({P_s} + 1){M_s}{N_c}}},
\end{equation}
where ${{\bf{\tilde w}}_R} = \frac{{{{\left[ {{{\bf{a}}^T}\left( {{\bf{\tilde p}}_{R,0}^U} \right)} \right]}^\dag }}}{{\sqrt {\left\| {{{\left[ {{{\bf{a}}^T}\left( {{\bf{\tilde p}}_{R,0}^U} \right)} \right]}^\dag }} \right\|_2^2} }} \in {\mathbb{C}^{{P_t}{Q_t} \times 1}}$ is the BF vector pointing at ${\bf{\tilde p}}_{R,0}^U$. Moreover, ${{{\bf{\tilde h}}}_R}$ can be expressed in a block matrix ${{{\bf{\tilde h}}}_R} = \left[ {{{{\bf{\bar h}}}_R},{\bf{\tilde h}}_R^d} \right]$, where ${\bf{\tilde h}}_R^d = {\left( {{{{\bf{\tilde w}}}_R}} \right)^H}{{\bf{\hat H}}^d} \in {\mathbb{C}^{1 \times {P_s}{M_s}{N_c}}}$. Reforming ${{\bf{\tilde h}}_R}$ into a new block matrix for range estimation, we obtain
\begin{equation}\label{equ:til_H_R}
	{{\bf{\tilde H}}_R} = \left[ {{{{\bf{\bar H}}}_R},{\bf{\tilde H}}_{R,1}^d, \cdots ,{\bf{\tilde H}}_{R,{P_s}}^d} \right] \in {\mathbb{C}^{{N_c} \times ({P_s} + 1){M_s}}},
\end{equation}
where ${\bf{\tilde H}}_{R,i}^d \in {\mathbb{C}^{{N_c} \times {M_s}}}$ is the matrix stacked by the D-CSI estimation of the $i$th OFDM symbol in each packet. 
Moreover, the $(n,m)$th element of ${\bf{\tilde H}}_{R,i}^d$ is 
\begin{equation}\label{equ:H_Rid_nm}
	\begin{array}{*{20}{l}}
		{{{\left[ {{\bf{\tilde H}}_{R,i}^d} \right]}_{n,m}} = {{\left( {{{{\bf{\tilde w}}}_R}} \right)}^H}{{\bf{h}}_{n,m}}\frac{1}{{1 + \psi _{n,m}^i}} + \tilde n_{n,m}^i}\\
		{ = \left( {\begin{array}{*{20}{l}}
					{\frac{1}{{1 + \psi _{n,m}^i}}\sqrt {P_t^U} {b_{C,0}}{{\chi }_{T,0}}{{\tilde \chi }_{R,0}}}\\
					{ \times {e^{j2\pi mT_s^p{{\tilde f}_{d,0,m}}}}{e^{ - j2\pi n\Delta f{{\tilde \tau }_{0,m}}}}}
			\end{array}} \right) + \tilde n_{n,m}^i},
	\end{array}
\end{equation}
where ${\tilde \chi _{R,0}} = {({{\bf{\tilde w}}_R})^H}{\bf{a}}({\bf{p}}_{R,0}^U)$, and $\tilde n_{n,m}^i$ is the transformed Gaussian noise with zero mean and variance $\sigma _N^2$.

Since ${\bf{\tilde H}}_{R,i}^d$ also has the same range steering vector form as \eqref{equ:a_r}, we can express the $m$th column of ${\bf{\tilde H}}_{R,i}^d$ as
\begin{equation}\label{equ:H_Rid_til}
	{\bf{\tilde h}}_{R,i,m}^d = {\left[ {{\bf{\tilde H}}_{R,i}^d} \right]_{:,m}} = {\tilde S_0} {\bf{\tilde a}}_{{\bf{r}},m}^i {\left[ {{{\bf{a}}_{\bf{f}}}} \right]_m} + {\left[ {{\bf{\tilde N}}_{R,i}^d} \right]_{_{:,m}}},
\end{equation}
where ${\tilde S_0} = {\sqrt {P_t^U} {b_{C,0}}{\chi _{T,0}}{{\tilde \chi }_{R,0}}}$, and ${\bf{\tilde N}}_{R,i}^d$ is the transformed noise matrix with ${\left[ {{\bf{\tilde N}}_{R,i}^d} \right]_{n,m}} = \tilde n_{n,m}^i$. Moreover, ${\bf{\tilde a}}_{{\bf{r}},m}^i$ is the range steering vector that can be distorted by the bit errors, which can be expressed as
\begin{equation}\label{equ:veca_rm}
	{\bf{\tilde a}}_{{\bf{r}},m}^i = [\frac{1}{{1 + \psi _{n,m}^i}}{e^{ - j2\pi n\Delta f\frac{{{{\tilde r}_m}}}{c}}}]{|_{n = 0,1,...,{N_c} - 1}} \in {\mathbb{C}^{{N_c} \times 1}},
\end{equation}

The MUSIC-based range estimation method exploits the autocorrelation of the signal matrix. The autocorrelation of ${{\bf{\tilde H}}_R}$ is expressed as
\begin{equation}\label{equ:R_r_til}
	\begin{array}{l}
		{{{\bf{\tilde R}}}_r} = \frac{1}{{({P_s} + 1){M_s}}}{{{\bf{\tilde H}}}_R}{( {{{{\bf{\tilde H}}}_R}} )^H}\\
		= \frac{1}{{({P_s} + 1){M_s}}}\left[ {{{{\bf{\bar H}}}_R}{{\left( {{{{\bf{\bar H}}}_R}} \right)}^H} + \sum\limits_{i = 1}^{{P_s}} {\sum\limits_{m = 0}^{{M_s} - 1} {{\bf{\tilde h}}_{R,i,m}^d{{\left( {{\bf{\tilde h}}_{R,i,m}^d} \right)}^H}} } } \right].
	\end{array}
\end{equation}
Since ${{{{\bf{\bar H}}}_R}{{\left( {{{{\bf{\bar H}}}_R}} \right)}^H}}$ is the autocorrelation of pure CSI and it can definitely contribute to the coherent signal processing, the key issue is analyzing the coherence of $\frac{1}{{({P_s} + 1){M_s}}}\sum\limits_{i = 1}^{{P_s}} {\sum\limits_{m = 0}^{{M_s} - 1} {{\bf{\tilde h}}_{R,i,m}^d{{\left( {{\bf{\tilde h}}_{R,i,m}^d} \right)}^H}} }$, which can be further expressed as 
\begin{equation}\label{equ:h_dim_expect}
	\begin{array}{l}
		\frac{1}{{({P_s} + 1){M_s}}}\sum\limits_{i = 1}^{{P_s}} {\sum\limits_{m = 0}^{{M_s} - 1} {{\bf{\tilde h}}_{R,i,m}^d{{\left( {{\bf{\tilde h}}_{R,i,m}^d} \right)}^H}} } \\
		\approx {E_{i,m}}\left\{ {{\bf{\tilde h}}_{R,i,m}^d{{\left( {{\bf{\tilde h}}_{R,i,m}^d} \right)}^H}} \right\}\\
		= {E_{i,m}}\left\{ \begin{array}{l}
			\left( {{{\tilde S}_0}{\bf{\tilde a}}_{{\bf{r}},m}^i{{\left[ {{{\bf{a}}_{\bf{f}}}} \right]}_m} + {\bf{\tilde n}}_{R,i,m}^d} \right)\\
			\times {\left( {{{\tilde S}_0}{\bf{\tilde a}}_{{\bf{r}},m}^i{{\left[ {{{\bf{a}}_{\bf{f}}}} \right]}_m} + {\bf{\tilde n}}_{R,i,m}^d} \right)^H}
		\end{array} \right\}\\
		= {E_{i,m}}\left\{ {{{\left| {{{\tilde S}_0}} \right|}^2}{\bf{\tilde a}}_{{\bf{r}},m}^i{{\left( {{\bf{\tilde a}}_{{\bf{r}},m}^i} \right)}^H} + {\bf{\tilde n}}_{R,i,m}^d{{\left( {{\bf{\tilde n}}_{R,i,m}^d} \right)}^H}} \right\}\\
		= {\left| {{{\tilde S}_0}} \right|^2}{E_{i,m}}\left\{ {{\bf{\tilde a}}_{{\bf{r}},m}^i{{\left( {{\bf{\tilde a}}_{{\bf{r}},m}^i} \right)}^H}} \right\} + \sigma _N^2{\bf{I}},
	\end{array}
\end{equation}
where ${\bf{\tilde n}}_{R,i,m}^d = {\left[ {{\bf{\tilde N}}_{R,i}^d} \right]_{_{:,m}}}$ is the transformed noise. Here, we focus on the autocorrelation of the distorted range steering vector as shown in \eqref{equ:veca_rm}, i.e., ${{{\bf{R}}_{{\bf{\tilde r}}}}} = {E_{i,m}}\left\{ {{\bf{\tilde a}}_{{\bf{r}},m}^i{{\left( {{\bf{\tilde a}}_{{\bf{r}},m}^i} \right)}^H}} \right\}$. The $(n_1, n_2)$th element of ${{{\bf{R}}_{{\bf{\tilde r}}}}}$ can be expressed as
\begin{equation}\label{equ:R_a_n1n2}
	\begin{array}{l}
		{\left[ {{{\bf{R}}_{{\bf{\tilde r}}}}} \right]_{{n_1},{n_2}}} = \\
		{e^{ - j2\pi {n^{'}}\Delta f{\tau _k}}}{E_{i,m}}\left\{ {\frac{1}{{\left( {1 + \psi _{{n_1},m}^i} \right){{\left( {1 + \psi _{{n_2},m}^i} \right)}^{*}}}}{e^{ - j2\pi {n^{'}}\Delta f{\delta _\tau }\left( m \right)}}} \right\},
	\end{array}
\end{equation}
where $n^{'} = n_1 - n_2$. Since ${\psi _{n,m}^i}$ is related to the symbol error, we can divide the symbol error situation of all the OFDM data symbols in $P_s$ packets into two cases to analyze the coherence of \eqref{equ:h_dim_expect}.
 
\begin{itemize}
	\item[1:] When the bit error rate is low enough and few or no symbol errors happen in the $m$th packet, then ${\left[ {{{\bf{R}}_{{\bf{\tilde a}}}}} \right]_{{n_1},{n_2}}}$ is equal to \eqref{equ:R_r_s_nm}. 
	
	\item[2:] When the bit error rate is high, and there is at least one symbol error in each pair of $(n_1,n_2)$th subcarriers in an OFDM symbol.
\end{itemize}

The MUSIC-based range estimation method can effectively suppress the noise-like TO-related phase shift effectively based on the D-CSI estimation for situation 1 as shown in Section~\ref{MUSIC_range_estimation}. 

Then, we focus on the situation 2 to analyze the influence of bit error rate (BER) on the coherence of signal autocorrelation for the MUSIC-based range estimation. In \textbf{Appendix \ref{appendix:Rr_0}}, we prove that ${\left[ {{{\bf{R}}_{{\bf{\tilde r}}}}} \right]_{{n_1},{n_2}}} = 0$ in situation 2. Therefore, BER will not deteriorate the coherence and sensing accuracy of the MUSIC-based range estimation, which will also be shown in Section~\ref{sec:Simulation}. Using the MUSIC-based range estimation method in Section~\ref{MUSIC_range_estimation} by replacing ${{\bf{R}}_r}$ with ${{{\bf{\tilde R}}}_r}$, we can obtain the range estimation of UE, denoted by $\tilde r_0$.

\subsubsection{Joint CSI and Data Signals-based Localization}

Based on the estimated AoA And range of UE obtained by the Joint CSI and data signal-based sensing scheme, we can further estimate the location of UE as
\begin{equation}\label{equ:location_JCDS}
	{{\bf{\tilde \Omega }}_0} = ({\tilde r_0}\sin {\tilde \theta _0}\cos {\tilde \varphi _0},{\tilde r_0}\sin {\tilde \theta _0}\sin {\tilde \varphi _0},{\tilde r_0}\cos {\tilde \theta _0}).
\end{equation}

\subsection{Complexity Analysis}

The proposed joint single-base localization and communication enhancement scheme contains two MUSIC-based procedures to estimate AoA and range of UE, and a CSI enhancement procedure. Since the CSI enhancement procedure is deeply coupled with the AoA estimation procedure as shown in \eqref{equ:H_u_bar}, the CSI enhancement procedure does not contribute extra complexity to the ISAC system. Therefore, the complexity of joint single-base localization and CSI enhancement in the preamble period as shown in Section~\ref{sec:JCAS_sensing} is ${\cal O}\{ {({P_t}{Q_t})^3} + {N_c}^3\}$.

The joint CSI and data signals-based localization adds $P_s M_s N_c$ D-CSI estimation based on OFDM data symbols to conduct MUSIC-based AoA and range estimation. Thus, the complexity of the joint CSI and data signal-based localization is approximately ${\cal O}\{ {({P_t}{Q_t})^3} + {N_c}^3 + {({P_t}{Q_t})^2}({P_s} + 1){M_s}{N_c} +  + {N_c}^2\left( {{P_s} + 1} \right){M_s}\} $.

\begin{figure}[!t]
	\centering
	\includegraphics[width=0.35\textheight]{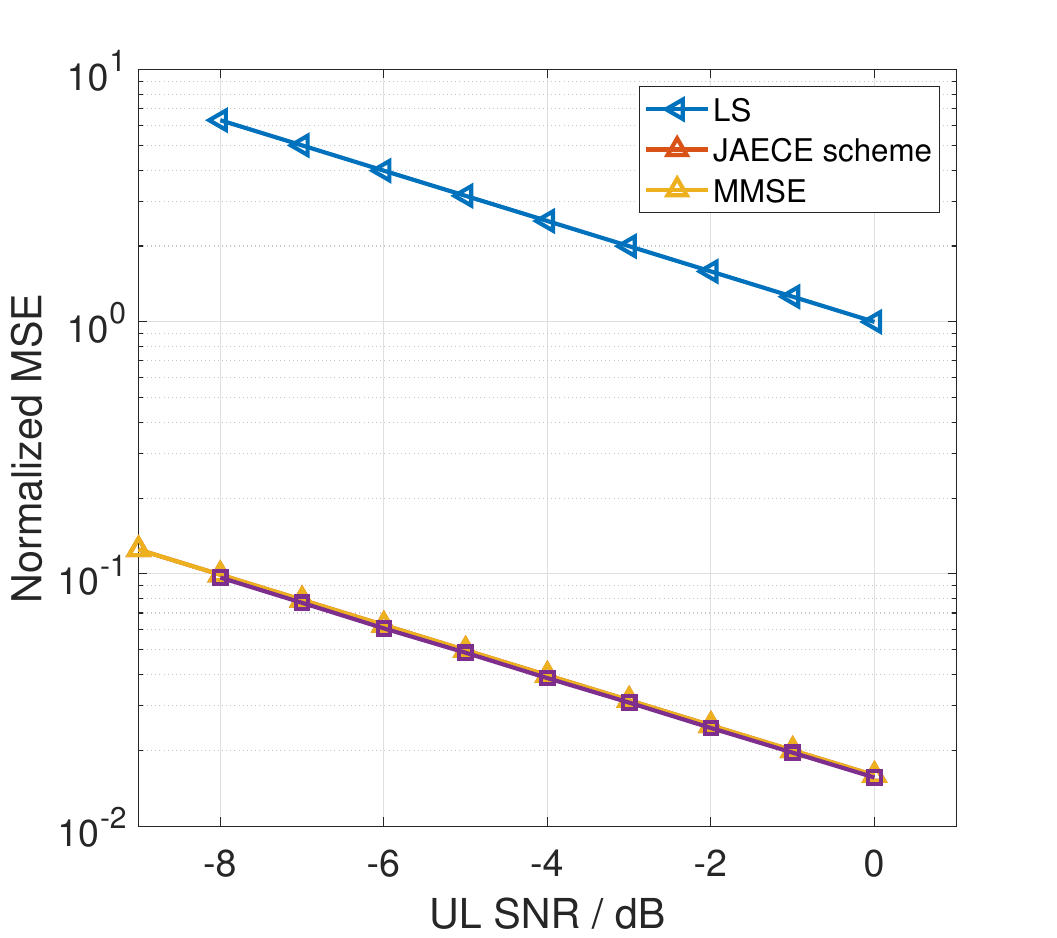}%
	\DeclareGraphicsExtensions.
	\caption{The normalized MSE for CSI estimation.}
	\label{fig: NMSE_CSI}
\end{figure}

\section{Simulation Results}\label{sec:Simulation}
In this section, we present the communication and sensing simulation results of the proposed joint single-base localization and CSI enhancement scheme and the joint CSI and data signals-based localization scheme.

\begin{figure*}[!t]
	\centering
	\subfigure[BERs under 4-QAM modulation.]{\includegraphics[width=0.35\textheight]
		{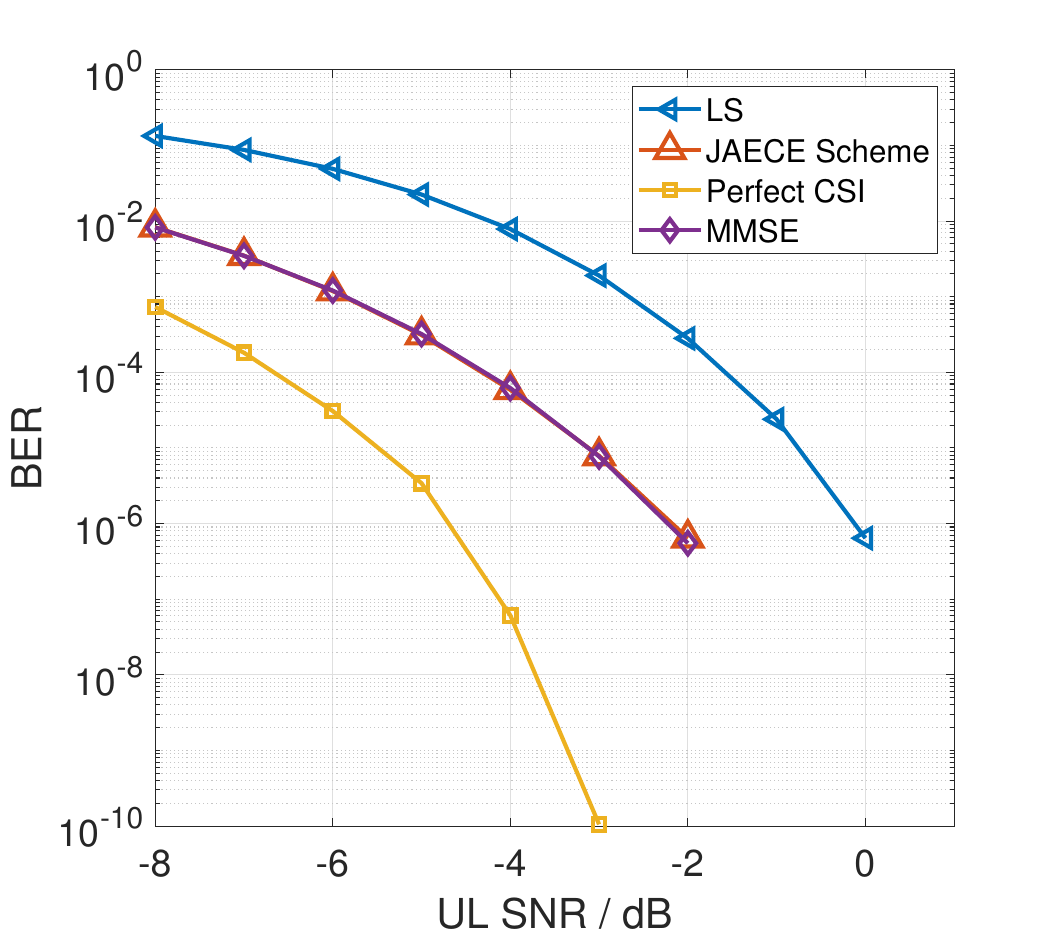}
		\label{fig:4QAM_BER}
	}
	\subfigure[BERs under 16-QAM modulation.]{\includegraphics[width=0.35\textheight]
		{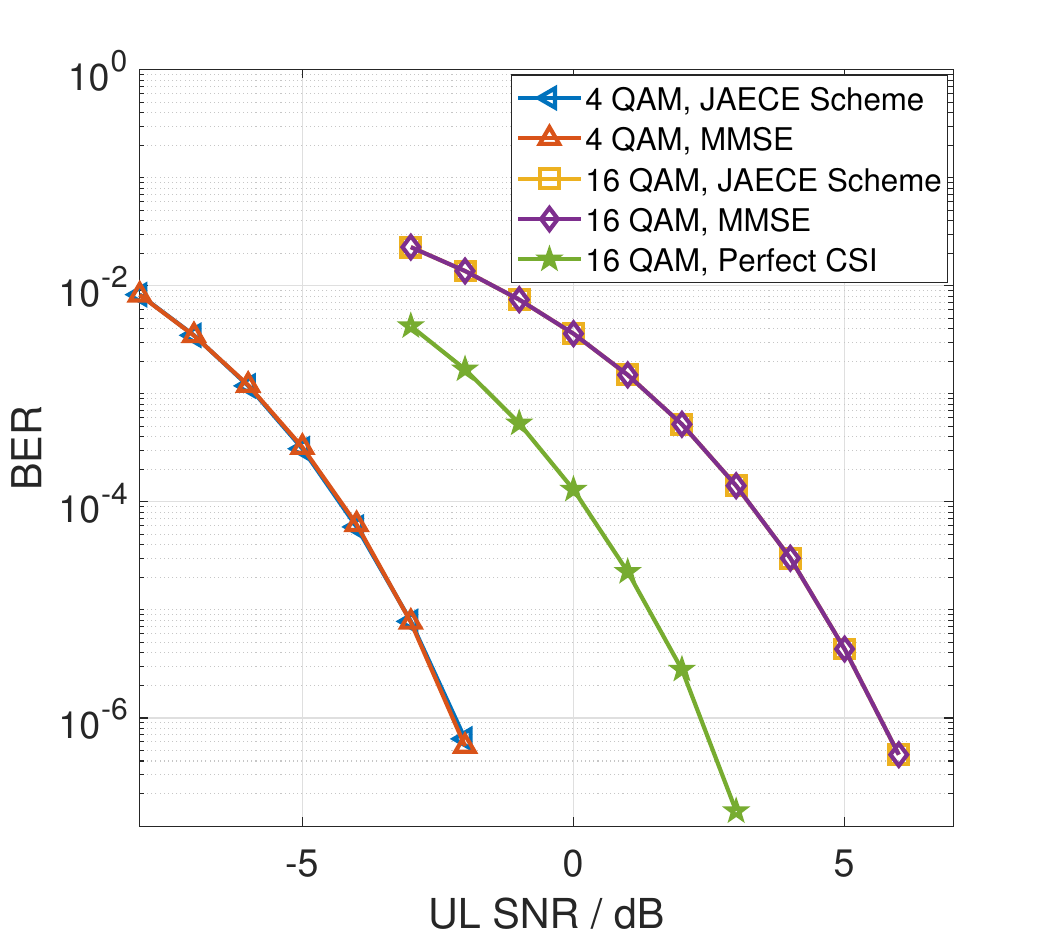}
		\label{fig:16QAM_BER}
	}
	\caption{The BERs using the JAECE scheme, LS, and MMSE methods under 4-QAM and 16-QAM modulation.}
	\label{fig:BER}
\end{figure*}

\begin{figure*}[!t]
	\centering
	\subfigure[AoA estimation MSEs of \textit{Schemes} 1 and 2 under various TOs, CFOs, and QAM orders when $M_s$ = 64.]{\includegraphics[width=0.36\textheight]
		{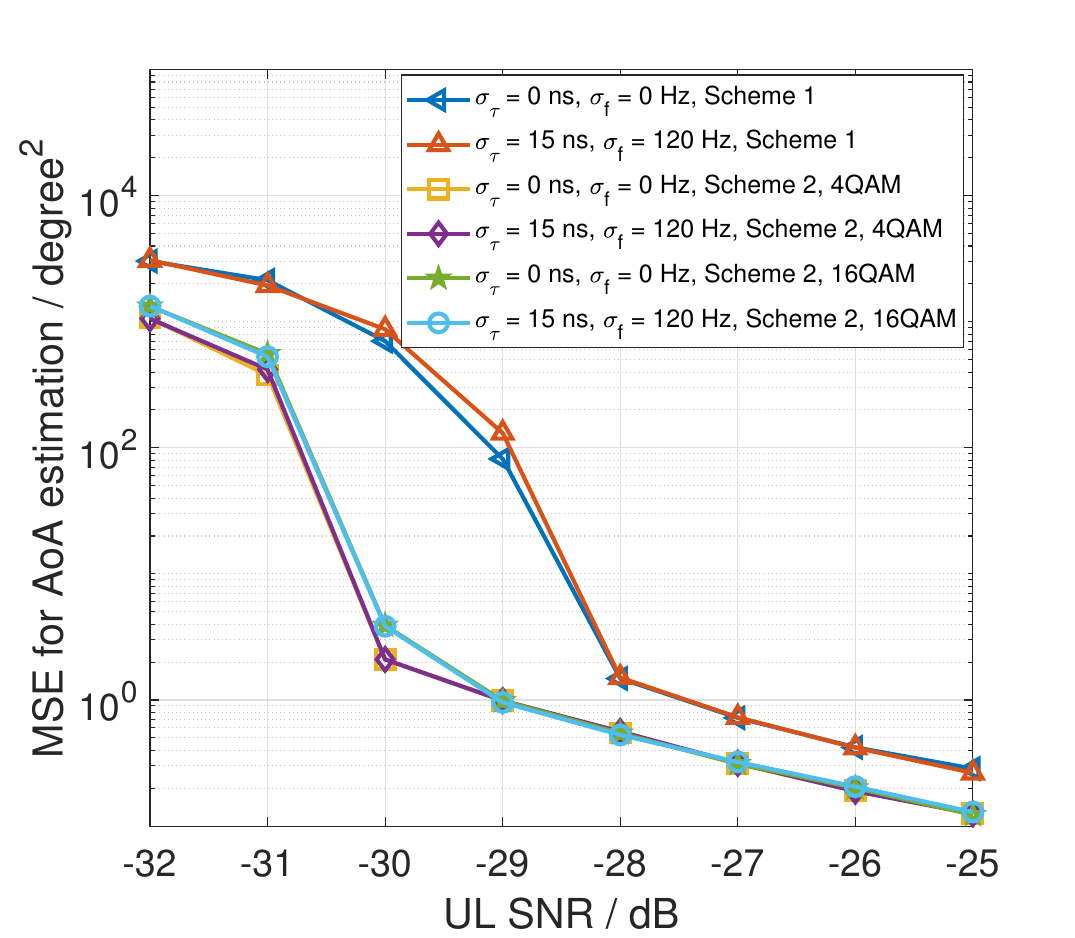}
		\label{fig: AoA_Ms64}
	}
	\subfigure[AoA estimation MSEs of \textit{Schemes} 1 and 2 under various TOs, CFOs, and $M_s$.]{\includegraphics[width=0.36\textheight]
		{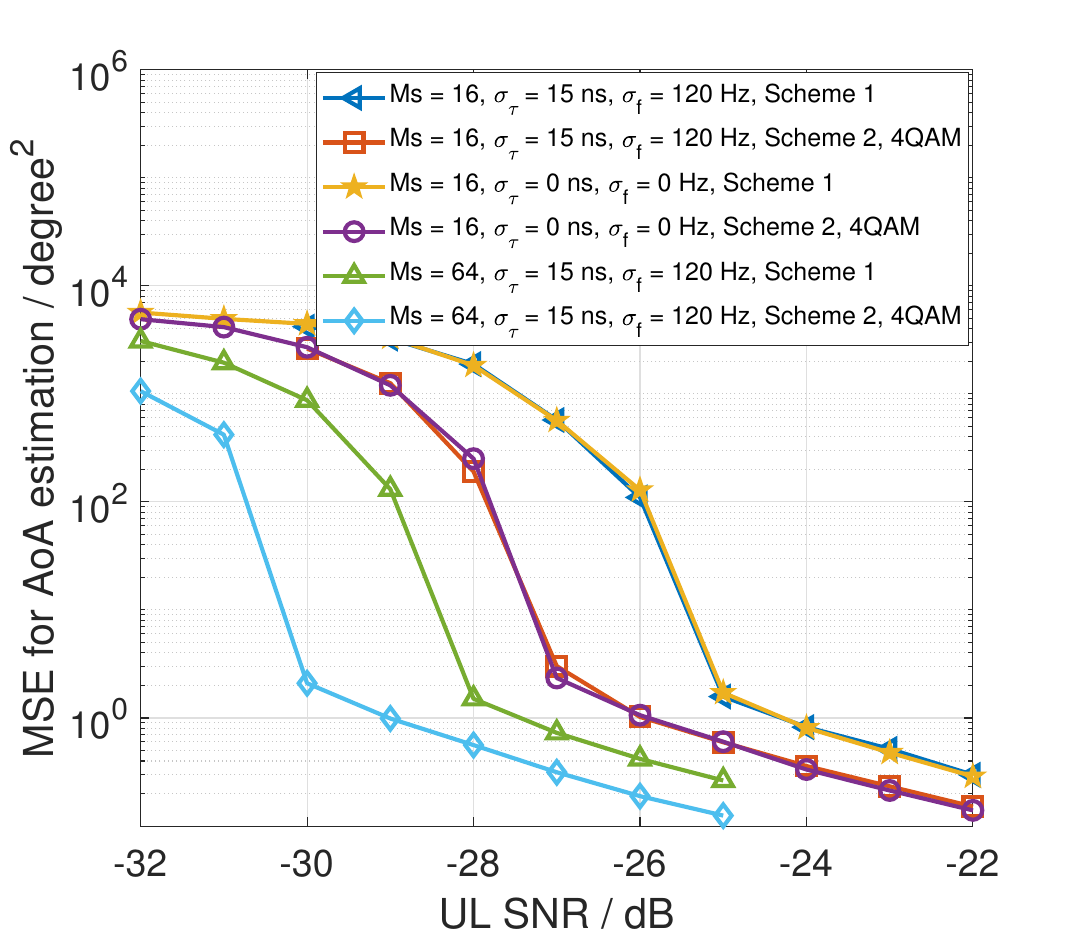}
		\label{fig: AoA_Ms_change}
	}
	\caption{The AoA estimation MSEs of \textit{Schemes} 1 and 2 under various TOs, CFOs, $M_s$, and QAM orders.}
	\label{fig:MSE_AoA}
\end{figure*}

\subsection{Simulation Setting}
The carrier frequency is set to 28 GHz, the antenna interval, $d_a$, is half of the wavelength, the sizes of antenna arrays of the BS and user are $P_t \times Q_t = 8 \times 8$ and $P_r \times Q_r = 1\times 1$, respectively. The subcarrier interval of UL preamble signal is $\Delta {f} =$ 480 kHz, the subcarrier number is $N_c =$ 256, and the bandwidth for JCAS is ${{B  =  }}{N_c}\Delta f = $122.88 MHz. 
The variance of the Gaussian noise is $\sigma_N^2 = kFTB = 4.9177\times10^{-12} $ W, where $k = 1.38 \times 10^{-23}$ J/K is the Boltzmann constant, $F = $ 10 is the noise factor, and $T = 290$ K is the standard temperature. The location of BS is (50, 4.75, 7) m. The location of UE is ($x_U$, 0, 2) m, where $x_U$ is uniformly distributed from 70 to 150 m. The velocity of UE is ($-$40, 0, 0) km/h, and the velocities of BS and the dumb scatterers are (0, 0, 0) m/s. 

The range and location estimation MSEs are defined as the mean values of all the square errors of the range and location estimation results under a certain set of simulation parameters. The RMSE is the square root of the MSE.

Based on the above locations and velocities of BS and user, the AoAs, AoDs, ranges, and Doppler shifts between UE and BS, and between the scatterers and BS can be derived to generate UL channel response matrix according to the models proposed in Section~\ref{subsec:JCAS_channel}. Further, BS can estimate the ranges and locations of UE according to Section~\ref{sec:JCAS_sensing}. Communication SNR is defined as the SNR of each antenna element of BS. According to \eqref{equ:h_c_U_bar}, the UL communication SNR is expressed as
\begin{equation}\label{equ:SNR}
	{\gamma _c} = \frac{{P_t^U\sum\limits_{k = 0}^{K - 1} {{{\left| {{b_{C,k}}\chi _{T,k}} \right|}^2}} }}{{\sigma _N^2}}.
\end{equation}

\subsection{Communication Performance of the JAECE Scheme}
This subsection presents the communication and AoA estimation performance of the proposed joint AoA estimation and CSI enhancement (JAECE) scheme in Section~\ref{sec:JCAS_sensing}. Moreover, we also show the AoA estimation performance of the joint CSI and data signal-based AoA estimation scheme proposed in Section~\ref{sec:JCDS_B_LOC}.

Fig.~\ref{fig: NMSE_CSI} shows the CSI estimation Normalized MSEs (NMSEs) of the proposed JAECE scheme compared with the LS and MMSE CSI estimators. It can be seen that the CSI estimation NMSEs of the JAECE scheme approach those of the MMSE estimation method and are about 16 dB lower than those of the LS method. 

Fig.~\ref{fig:BER} presents the BER performance of the enhanced CSI estimation of the proposed JAECE scheme compared with those using perfect CSI estimation and the CSIs estimated by the LS and MMSE methods under 4-QAM and 16-QAM modulation, respectively. 
Fig.~\ref{fig:4QAM_BER} shows the BER performance under 4-QAM modulation. The BERs of demodulation using the JAECE scheme are generally the same as those using the MMSE method. Moreover, the JAECE scheme requires about 3 dB SNR lower than the LS method to achieve the same BER. 
Fig.~\ref{fig:16QAM_BER} shows the BERs of demodulation using the proposed JAECE scheme and the MMSE method under 16-QAM modulation. It can be seen that the BERs of demodulation using the JAECE scheme still approach those using the MMSE method. It requires about 3 dB higher SNR for the JAECE scheme to achieve the same BER performance as the MMSE method.
%
%

\subsection{Sensing Performance}

We predefine \textit{schemes} 1 and 2 for the simplicity of demonstration. \textit{Scheme} 1 refers to the single-base localization scheme presented in Section~\ref{sec:JCAS_sensing}, while \textit{scheme} 2 refers to the joint CSI and data signals-based localization scheme proposed in Section~\ref{sec:Joint_CSI_data_localization} when $P_s$ = 1.

Figs.~\ref{fig:MSE_AoA} presents the AoA estimation MSEs of \textit{schemes} 1 and 2 under various $\sigma_\tau$, $\sigma_f$, $M_s$, and QAM orders. As SNR increases, the AoA estimation MSE decreases due to the accumulation of sensing energy. From Fig.~\ref{fig: AoA_Ms64}, we can see that given the same $M_s$ and QAM order, the AoA estimation MSEs of the same scheme are not affected prominently by the change of $\sigma_\tau$ and $\sigma_f$, which verifies the corresponding statement in Section~\ref{sec:AoA_estimation_CSI}. The AoA estimation MSEs of \textit{scheme} 2 are lower than those of \textit{scheme} 1 for both situations when using 4 or 16 QAM modulation. This is because \textit{scheme} can accumulate the energy of both data signals and CSIs coherently to enhance the AoA estimation as proved in \eqref{equ:R_H_tilde} in Section~\ref{sec:JCDS_AoA}. Moreover, given the same $M_s$, we can see that the AoA estimation MSEs of \textit{scheme} 2 under 16 QAM are larger than those under 4 QAM. This is because the larger QAM order leads to larger equivalent noise as shown in \eqref{equ:h_nm_i}. Fig.~\ref{fig: AoA_Ms_change} shows the AoA estimation MSEs of \textit{schemes} 1 and 2 under various $\sigma_\tau$, $\sigma_f$ and $M_s$. We can see that $M_s$ influences the AoA estimation MSEs concretely. Given the same used scheme and QAM order, more OFDM packets used for AoA estimation results in lower AoA estimation MSEs. This is because the larger $M_s$ contributes more coherent sensing energy as shown in \eqref{equ:R_H_tilde}. 

\begin{figure*}[!t]
	\centering
	\subfigure[The MSEs of range estimation under various TOs and CFOs when $M_s$ = 64.]{\includegraphics[width=0.34\textheight]
		{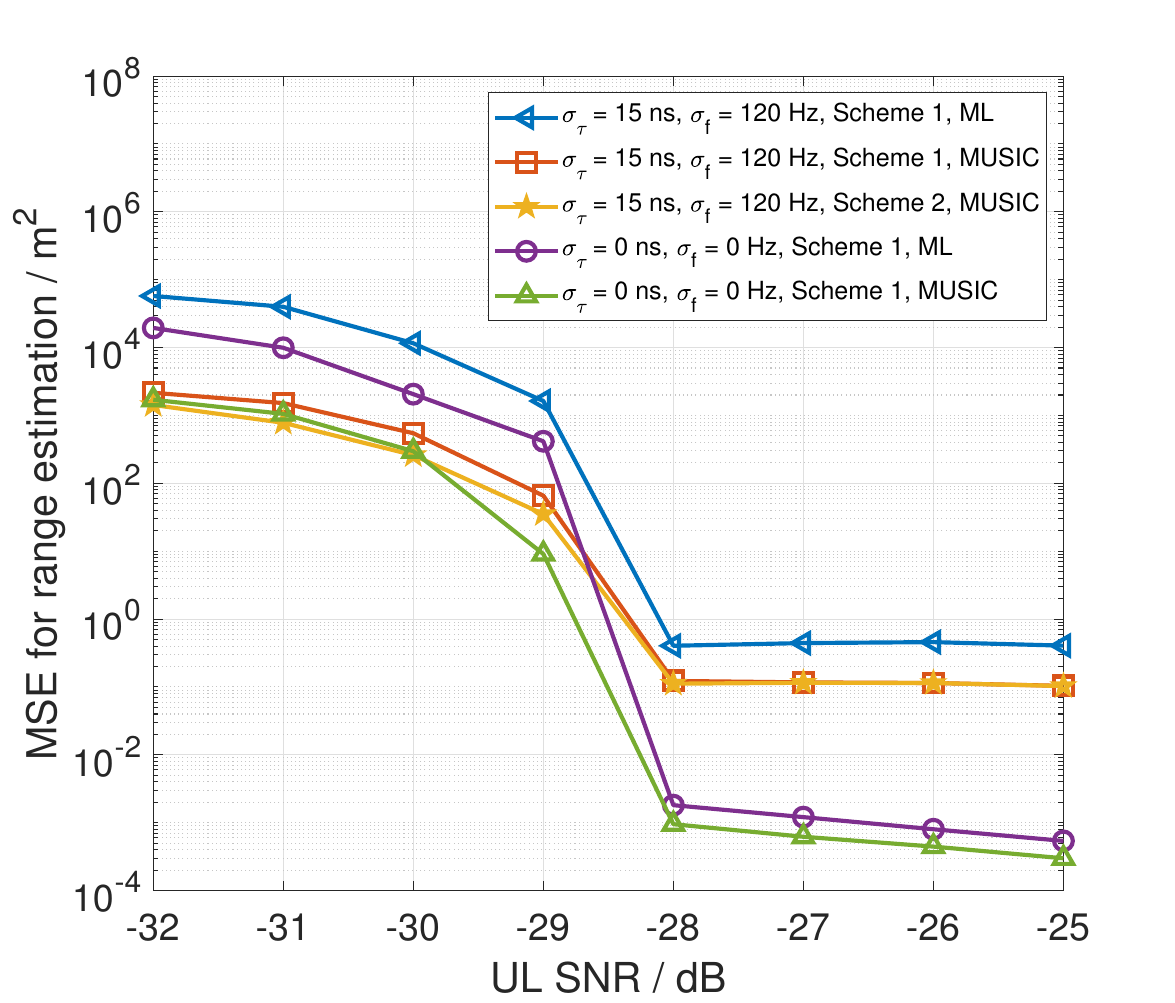}
		\label{fig: MSE_range_Ms64}
	}
	\subfigure[The MSEs of localization under various TOs and CFOs when $M_s$ = 64.]{\includegraphics[width=0.34\textheight]
		{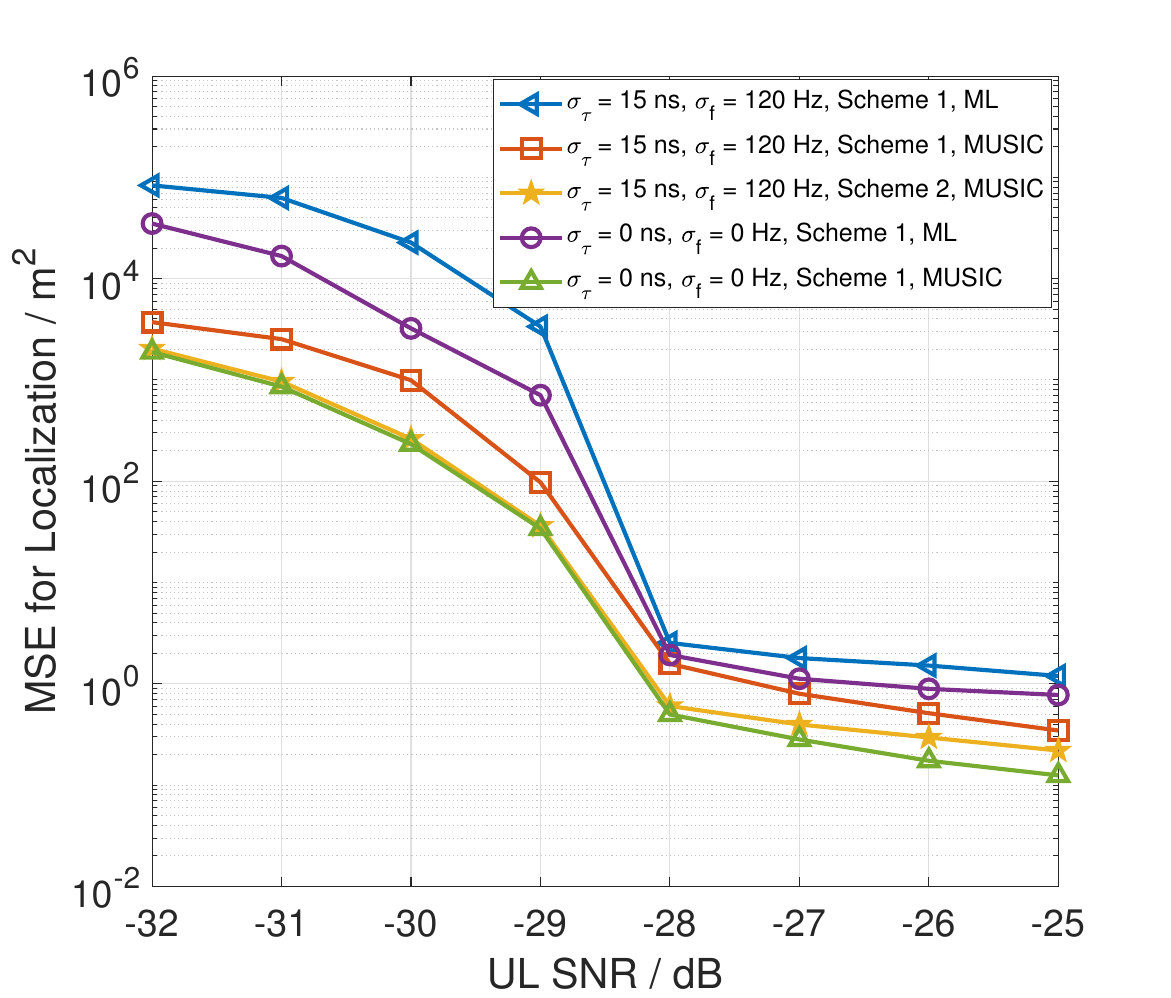}
		\label{fig: MSE_Loc_Ms64}
	}\\
	\subfigure[The MSEs of range estimation under various $M_s$ when $\sigma_{\tau}$ = 15 ns and $\sigma_f$ = 120 Hz.]{\includegraphics[width=0.34\textheight]
		{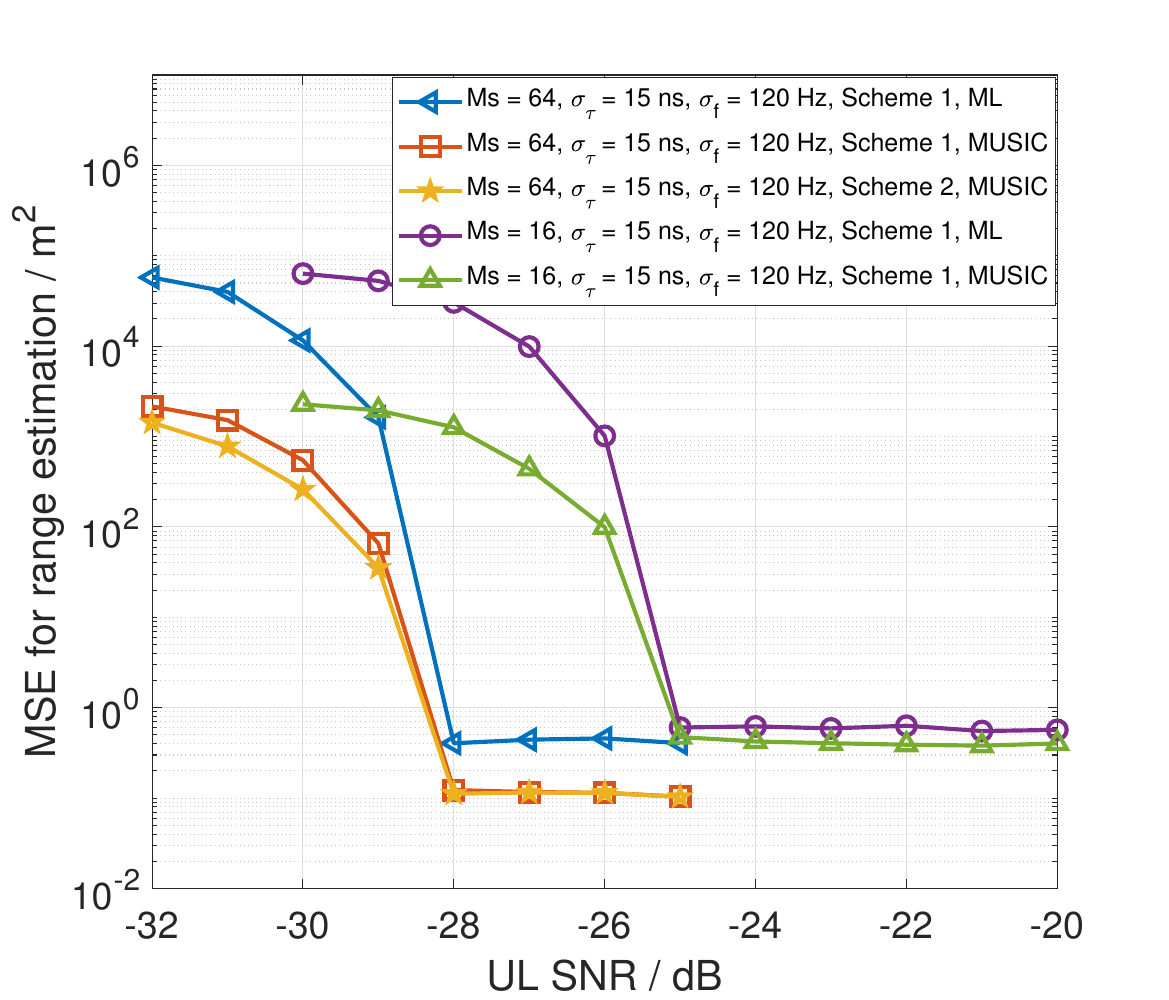}
		\label{fig: MSE_range_Ms_change}
	}
	\subfigure[The MSEs of localization under various $M_s$ when $\sigma_{\tau}$ = 15 ns and $\sigma_f$ = 120 Hz.]{\includegraphics[width=0.34\textheight]
		{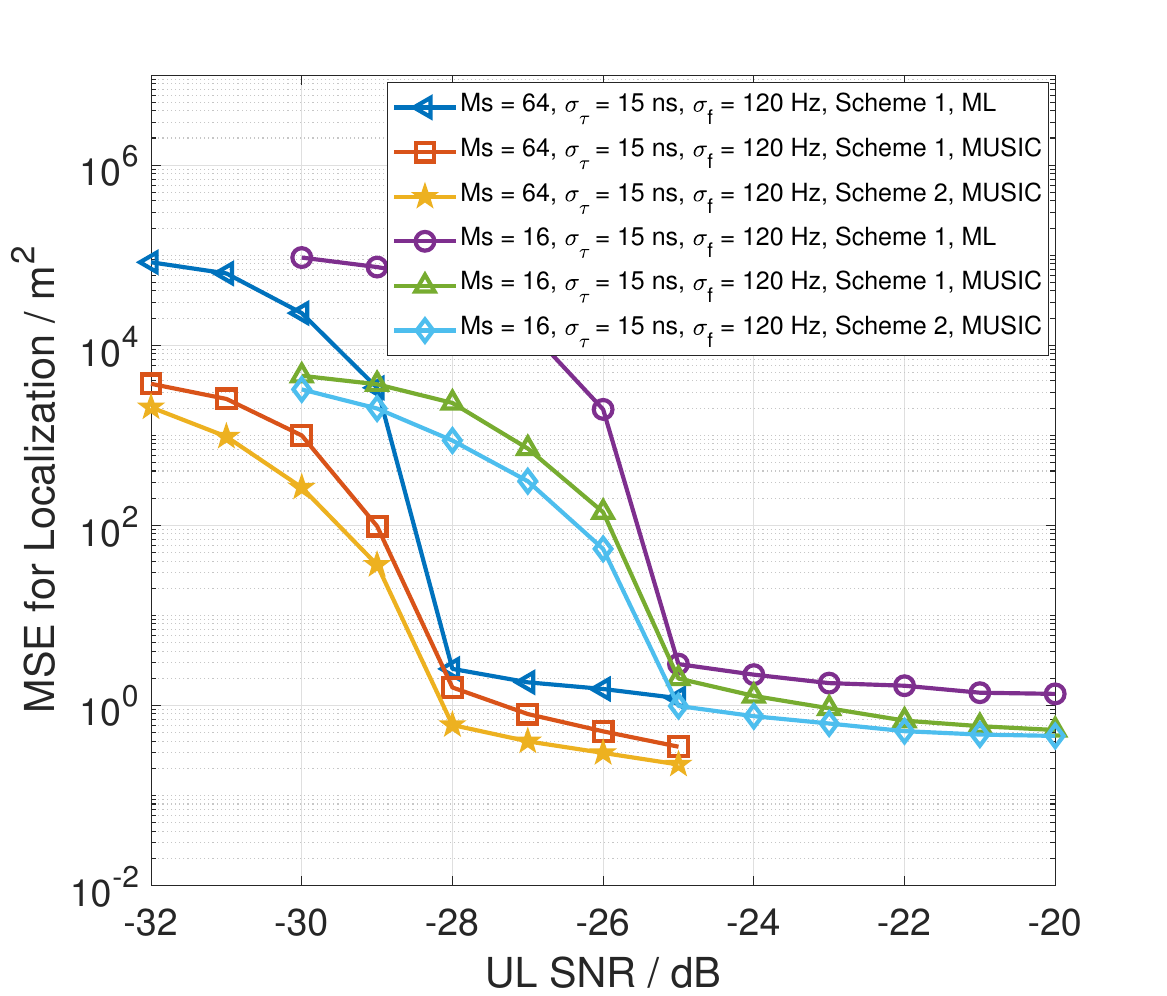}
		\label{fig: MSE_loc_Ms_change}
	}
	\caption{The MSEs of range estimation and localization of \textit{Schemes} 1 and 2 under various TOs, CFOs, and $M_s$.}
	\label{fig:MSE_Range_location}
\end{figure*}

Fig.~\ref{fig:MSE_Range_location} shows the MSEs of range estimation and localization of UE using \textit{schemes} 1 and 2 under various $\sigma_\tau$, $\sigma_f$ and $M_s$ when using 4-QAM modulation. The maximum-likelihood (ML)-based range estimation method~\cite{Zhang2019JCRS} can be used to estimate the range and localize the targets when AoA is estimated. Figs.~\ref{fig: MSE_range_Ms64} and \ref{fig: MSE_Loc_Ms64} present the range and location estimation MSEs of UE under various TOs and CFOs when $M_s$ = 64. When $\sigma_\tau$ = 0 ns and $\sigma_f$ = 0 Hz, it refers to the ideal situation where TO and CFO do not exist.

Fig.~\ref{fig: MSE_range_Ms64} shows that when using the \textit{scheme} 1, the MUSIC-based range estimation method can achieve about 6 dB lower range estimation MSEs compared with the ML-based range estimation given the same $\sigma_\tau$ and $\sigma_f$. This verifies that the MUSIC-based method can suppress the TO-related noise-like terms exploiting the randomness of TO in various OFDM packets as shown in \textbf{Proposition~\ref{proposition_MUSIC_range}}. Moreover, given the same $\sigma_\tau$ and $\sigma_f$, The range estimation MSEs of \textit{scheme} 2 in the low SNR regime are lower than those of using \textit{scheme} 1. This is because \textit{scheme} 2 can accumulate more coherent energy for sensing as shown in \eqref{equ:h_dim_expect}. 

Fig.~\ref{fig: MSE_Loc_Ms64} presents that the localization MSEs decrease as SNR increases since the AoA and range estimation MSEs both decrease as SNR increases according to Figs.~\ref{fig:MSE_AoA} and \ref{fig: MSE_range_Ms64}. Given the same $\sigma_\tau$ and $\sigma_f$ for \textit{scheme} 1, we can see that the localization MSEs of using the MUSIC-based range estimation method are about 5 dB lower than those using the ML-based range estimation method. This is because the TO-related range ambiguity can be better suppressed than the conventional ML-based sensing method according to \ref{fig: MSE_range_Ms64}. Moreover, when \textit{scheme} 2 and the MUSIC-based sensing method are used, the corresponding localization MSEs are about 8 dB lower than those with \textit{scheme} 1 and the ML-based sensing method. This is because \textit{scheme} 2 can further achieve better AoA estimation accuracy than \textit{scheme} 1 as shown in Fig.~\ref{fig:MSE_AoA}. We can see that the combination of \textit{scheme} 2 and the MUSIC-based sensing method can achieve decimeter-level single-base localization, while the conventional ML-based sensing method with scheme 1 can realize millimeter-level single-base localization.

Figs.~\ref{fig: MSE_range_Ms_change} and \ref{fig: MSE_loc_Ms_change} show the range and location estimation MSEs of UE under various $M_s$ when $\sigma_{\tau}$ = 15 ns and $\sigma_f$ = 120 Hz. 

Fig.~\ref{fig: MSE_range_Ms_change} shows that the larger $M_s$ is, the smaller the MSEs of range estimation are. This is because the larger $M_s$ contributes more coherent energy for range sensing as shown in \eqref{equ:R_r_til} and \eqref{equ:h_dim_expect}. When $M_s$ = 64, the required SNR to achieve the minimum range estimation MSE is about 3 dB lower than that for $M_s$ = 16. Moreover, we can see that when $M_s$ decreases, the range estimation MSE gap between the ML-based and MUSIC-based sensing methods becomes smaller. This is because more packets offer more terms to suppress the TO-related phase noise as shown in \eqref{equ:R_r_s_nm} in \textbf{Appendix~\ref{appendix:MUSIC}}.

From Fig.~\ref{fig: MSE_loc_Ms_change}, we can see that given the same scheme and the range estimation method, the smaller $M_s$ leads to the larger localization MSEs, which results from the larger AoA and range estimation MSEs according to Figs.~\ref{fig:MSE_AoA} and \ref{fig: MSE_range_Ms_change}, respectively. Moreover, it is shown that the larger $M_s$ is, the larger the localization MSE gap between using the \textit{schemes} 1 and 2 is. This is because more packets offer more terms for suppressing the TO-related phase noise as shown in \eqref{equ:R_r_s_nm} in \textbf{Appendix~\ref{appendix:MUSIC}} and for accumulating the coherent energy for AoA estimation as shown in \eqref{equ:R_H_tilde}. Moreover, we can see that \textit{scheme} 2 and the MUSIC-based sensing method can achieve decimeter-level single-base localization compared with the millimeter-level single-base localization achieved by the conventional ML-based sensing combined with \textit{scheme} 1.

\section{Conclusion}\label{sec:conclusion}

In this paper, we propose a joint single-base localization and communication enhancement scheme for the UL ISAC system. We first propose the JAECE scheme that integrates the CSI enhancement procedure into the MUSIC-based AoA estimation and imposes no additional complexity on the ISAC system. We further prove that the MUSIC-based range estimation method can suppress the time-varying TO-related phase terms by exploiting the averaging effects of the noise-like phase terms on the timeframe direction. Finally, we propose a joint CSI and data signals-based localization scheme that can coherently exploit the data signals with the CSI signals to improve the AoA and range estimation for localizing UE. Simulation results show that the BER performance of demodulation using the proposed JAECE scheme is equivalent to that using the MMSE method, and the localization MSEs using the proposed joint CSI and data signals-based localization scheme are about 8 dB lower than those of the ML-based benchmark method in the high SNR regime.

\begin{appendices} 

\section{Solution to \eqref{equ:optimize_H}} \label{appendix:optimize_H}

Denote the objective of \eqref{equ:optimize_H} as
\begin{equation}\label{equ:J}
	\begin{array}{l}
		J = E\left\| {{\bf{B\bar H}} - {\bf{H}}} \right\|_F^2\\
		= E\{\rm Tr[\left( {{\bf{B\bar H}} - {\bf{H}}} \right){\left( {{\bf{B\bar H}} - {\bf{H}}} \right)^H}]\} \\
		= E\{\rm Tr\left( {{\bf{B\bar H}}{{{\bf{\bar H}}}^H}{{\bf{B}}^H} + {\bf{H}}{{\bf{H}}^H} - {\bf{B\bar H}}{{\bf{H}}^H} - {\bf{H}}{{{\bf{\bar H}}}^H}{{\bf{B}}^H}} \right)\} \\
		= \rm Tr\left( {{\bf{B}}{{\bf{R}}_{{\bf{\bar H}}}}{{\bf{B}}^H} + {{\bf{R}}_{\bf{H}}} - {\bf{B}}{{\bf{R}}_{\bf{H}}} - {{\bf{R}}_{\bf{H}}}{{\bf{B}}^H}} \right)
	\end{array}.
\end{equation}

Since \eqref{equ:J} is a convex problem, the optimal solution to \eqref{equ:optimize_H}, denoted by ${{{\bf{B}}_{opt}}}$, should satisfy
\begin{equation}\label{equ:J_opt}
	{\left. {\frac{{\partial J}}{{\partial {\bf{B}}}} = {\bf{0}}} \right|_{{\bf{B}} = {{\bf{B}}_{opt}}}}.
\end{equation}

Combining \eqref{equ:J} and \eqref{equ:J_opt}, we obtain
\begin{equation}\label{equ:J_opt_derive}
	\begin{array}{l}
		\frac{{\partial J}}{{\partial {\bf{B}}}} = {\bf{B}}\left( {{{\bf{R}}_{{\bf{\bar H}}}}^H + {{\bf{R}}_{{\bf{\bar H}}}}} \right) - {{\bf{R}}_{\bf{H}}}^H - {{\bf{R}}_{\bf{H}}}\\
		= 2{\bf{B}}{{\bf{R}}_{{\bf{\bar H}}}} - 2{{\bf{R}}_{\bf{H}}}
	\end{array}.
\end{equation}

Finally, by combining \eqref{equ:J_opt} and \eqref{equ:J_opt_derive}, we derive the optimal solution to \eqref{equ:optimize_H} as shown in \eqref{equ:optimal_B}.

\section{Proof of {Proposition~\ref{proposition_MUSIC_range}}}~\label{appendix:MUSIC}

The autocorrelation of ${{{{\bf{\bar H}}}_R}}$ can be further expressed as
\begin{equation}\label{equ:R_r_2}
	\begin{array}{l}
		{{\bf{R}}_r} = \frac{1}{{{M_s}}}\sum\limits_{m = 0}^{{M_s} - 1} {{{\left[ {{{{\bf{\bar H}}}_R}} \right]}_{:,m}}{{\left( {{{\left[ {{{{\bf{\bar H}}}_R}} \right]}_{:,m}}} \right)}^H}} \\
		= \frac{1}{{{M_s}}}\sum\limits_{m = 0}^{{M_s} - 1} {\left[ \begin{array}{l}
				{\left| {{S_0}{{\left[ {{{\bf{a}}_{\bf{f}}}} \right]}_m}} \right|^2}{{\bf{a}}_{{\bf{r}},m}}{\left( {{{\bf{a}}_{{\bf{r}},m}}} \right)^H} + {{{\bf{\bar n}}}_{u,m}}{\left( {{{{\bf{\bar n}}}_{u,m}}} \right)^H}\\
				+ 2{\mathop{\rm Re}\nolimits} \left( {{S_0}{{\left[ {{{\bf{a}}_{\bf{f}}}} \right]}_m}{{\bf{a}}_{{\bf{r}},m}}{{\left( {{{{\bf{\bar n}}}_{u,m}}} \right)}^H}} \right)
			\end{array} \right]},
	\end{array}
\end{equation}
where ${{{\bf{\bar n}}}_{u,m}} = {\left[ {{{{\bf{\bar N}}}_R}} \right]_{_{:,m}}}$. Since ${{{\bf{\bar n}}}_{u,m}}$ is irrelevant to ${{{\bf{a}}_{{\bf{r}},m}}}$, ${{\bf{R}}_r}$ can be further expressed as
\begin{equation}\label{equ:R_r_3}
	{{\bf{R}}_r} = \frac{1}{{{M_s}}}\sum\limits_{m = 0}^{{M_s} - 1} {\left[ {{{\left| {{S_0}{{\left[ {{{\bf{a}}_{\bf{f}}}} \right]}_m}} \right|}^2}{{\bf{a}}_{{\bf{r}},m}}{{\left( {{{\bf{a}}_{{\bf{r}},m}}} \right)}^H} + {{{\bf{\bar n}}}_{u,m}}{{\left( {{{{\bf{\bar n}}}_{u,m}}} \right)}^H}} \right]}.
\end{equation}

We focus on the signal part of ${{\bf{R}}_r}$ to analyze the suppression of TO-related phase-shift terms. The autocorrelation contributed by the useful signal can be expressed by
\begin{equation}\label{equ:R_r_s}
	{{\bf{R}}_{r,s}} = \frac{{{{\left| {{S_0}} \right|}^2}}}{{{M_s}}}\sum\limits_{m = 0}^{{M_s} - 1} {\left[ {{{\bf{a}}_{{\bf{r}},m}}{{\left( {{{\bf{a}}_{{\bf{r}},m}}} \right)}^H}} \right]}.
\end{equation}

By substituting \eqref{equ:a_r} into \eqref{equ:R_r_s}, the $(n_1,n_2)$th element of ${{\bf{R}}_{r,s}}$ can be expressed as
\begin{equation}\label{equ:R_r_s_nm}
	{\left[ {{{\bf{R}}_{r,s}}} \right]_{{n_1},{n_2}}} = \frac{{{{\left| {{S_0}} \right|}^2}{e^{ - j2\pi {n^{'}}\Delta f{\tau _k}}}}}{{{M_s}}}\left( {\sum\limits_{m = 0}^{{M_s} - 1} {{e^{ - j2\pi {n^{'}}\Delta f{\delta _\tau }\left( m \right)}}} } \right),
\end{equation}
where $n^{'} = n_1-n_2$. Since ${\delta _\tau }\left( m \right)$ follows zero-mean stochastic distribution, ${\delta _\tau }\left( m \right)$ can be reduced to nanosecond-level~\cite{Zhang2022ISAC}, and $\Delta f$ is from 15 kHz to 480 kHz, $\frac{1}{{{M_s}}}\left[ {\sum\limits_{m = 0}^{{M_s} - 1} {{e^{ - j2\pi {n^{'}}\Delta f{\delta _\tau }\left( m \right)}}} } \right]$ approaching 1 is satisfied for most $n^{'}$ when ${M_s} \to \infty$.

Therefore, the MUSIC-based range estimation method can suppress the TO-related phase shift.

\section{Proof of Proposition~\ref{proposition_AoA}}\label{appendix:AoA_enhancement}

Since ${{{\bf{\bar H}}}_u} {{\left( {{{{\bf{\bar H}}}_u}} \right)}^H}$ is the P-CSI related term and will definitely contribute to the coherent energy for sensing, ${\sum\limits_{n,m,i} {{\bf{\hat h}}_{n,m}^i{{( {{\bf{\hat h}}_{n,m}^i} )}^H}} }$ is the key part for analyzing whether the joint CSI and data signals can be used to improve the sensing performance. According to \eqref{equ:h_nm_i}, we obtain
\begin{equation}\label{equ:h_h_auto}
	\begin{array}{l}
		\sum\limits_{n,m,i}^{} { {\bf{\hat h}}_{n,m}^i{{( {{\bf{\hat h}}_{n,m}^i} )}^H}} \\
		= \sum\limits_{n,m,i}^{} {\left( {{{\bf{h}}_{n,m}}\frac{1}{{1 + \psi _{n,m}^i}} + {\bf{\hat n}}_{n,m}^i} \right){{\left( {{{\bf{h}}_{n,m}}\frac{1}{{1 + \psi _{n,m}^i}} + {\bf{\hat n}}_{n,m}^i} \right)}^H}} \\
		= \sum\limits_{n,m,i}^{} {\left( {{{\left| {\frac{1}{{1 + \psi _{n,m}^i}}} \right|}^2}{{\bf{h}}_{n,m}}{{\left( {{{\bf{h}}_{n,m}}} \right)}^H} + {\bf{\hat n}}_{n,m}^i{{\left( {{\bf{\hat n}}_{n,m}^i} \right)}^H}} \right)}.
	\end{array}
\end{equation}

Note that in \eqref{equ:h_h_auto}, we omit the cross-multiplication of ${\bf{\hat n}}_{n,m}^i$ and ${\bf{\hat h}}_{n,m}^i$ because its expectation is 0 in \eqref{equ:R_H_tilde}. 

Since $\psi_{n,m}^i = \frac{{e_{n,m}^i}}{{d_{n,m}^i}}$ in \eqref{equ:h_h_auto} is not related to the antenna-relevant indexes, i.e., $p$ and $q$, the sensing energy for AoA estimation can still be accumulated coherently. However, as the QAM-order used for modulation becomes higher, the BER increases and ${{{\left| {\frac{1}{{1 + \psi _{n,m}^i}}} \right|}^2}}$ decreases, resulting in smaller processing SNR for AoA estimation. 

\section{Proof of ${\left[ {{{\bf{R}}_{{\bf{\tilde r}}}}} \right]_{{n_1},{n_2}}} = 0$ in situation 2}\label{appendix:Rr_0}

Here, ${\left[ {{{\bf{R}}_{{\bf{\tilde r}}}}} \right]_{{n_1},{n_2}}}$ can be rewritten as
\begin{equation}\label{equ:R_r_til_nm}
	\begin{array}{l}
		{\left[ {{{\bf{R}}_{{\bf{\tilde r}}}}} \right]_{{n_1},{n_2}}}\\
		= {e^{ - j2\pi {n^{'}}\Delta f{\tau _k}}}{E_{i,m}}\left\{ {\frac{{d_{{n_1},m}^i}}{{\hat d_{{n_1},m}^i}}\frac{{{{\left( {d_{{n_2},m}^i} \right)}^{*}}}}{{{{\left( {\hat d_{{n_2},m}^i} \right)}^{*}}}}{e^{ - j2\pi {n^{'}}\Delta f{\delta _\tau }\left( m \right)}}} \right\},
	\end{array}
\end{equation}
where ${d_{{n_1},m}^i}$ and ${d_{{n_2},m}^i}$ are the actual data symbols, and ${\hat d_{{n_1},m}^i}$ and ${\hat d_{{n_2},m}^i}$ are the corresponding decoded symbols. Since the transmitted data symbols are irrelevant random symbols, and the symbol errors are caused by the i.i.d. Gaussian noise, we obtain ${E_{i,m}}\left\{ {d_{{n_1},m}^i{{\left( {d_{{n_2},m}^i} \right)}^*}} \right\} = 0$. Therefore, we can further derive
\begin{equation}\label{equ:R_r_n1n2}
	\begin{array}{l}
		{\left[ {{{\bf{R}}_{{\bf{\tilde r}}}}} \right]_{{n_1},{n_2}}}\\
		= {e^{ - j2\pi {n^{'}}\Delta f{\tau _k}}}{E_{i,m}}\{ {d_{{n_1},m}^i{{( {d_{{n_2},m}^i} )}^{*}}} \}{E_{i,m}}\left\{ {\frac{{{e^{ - j2\pi {n^{'}}\Delta f{\delta _\tau }\left( m \right)}}}}{{\hat d_{{n_1},m}^i{{\left( {\hat d_{{n_2},m}^i} \right)}^{*}}}}} \right\}\\
		= 0
	\end{array},
\end{equation}

\end{appendices}



%

{\small
	\bibliographystyle{IEEEtran}
	\bibliography{reference}

\begin{thebibliography}{10}
\providecommand{\url}[1]{#1}
\csname url@samestyle\endcsname
\providecommand{\newblock}{\relax}
\providecommand{\bibinfo}[2]{#2}
\providecommand{\BIBentrySTDinterwordspacing}{\spaceskip=0pt\relax}
\providecommand{\BIBentryALTinterwordstretchfactor}{4}
\providecommand{\BIBentryALTinterwordspacing}{\spaceskip=\fontdimen2\font plus
\BIBentryALTinterwordstretchfactor\fontdimen3\font minus
  \fontdimen4\font\relax}
\providecommand{\BIBforeignlanguage}[2]{{%
\expandafter\ifx\csname l@#1\endcsname\relax
\typeout{** WARNING: IEEEtran.bst: No hyphenation pattern has been}%
\typeout{** loaded for the language `#1'. Using the pattern for}%
\typeout{** the default language instead.}%
\else
\language=\csname l@#1\endcsname
\fi
#2}}
\providecommand{\BIBdecl}{\relax}
\BIBdecl

\bibitem{Feng2021JCSC}
Z.~Feng, Z.~Wei, X.~Chen, H.~Yang, Q.~Zhang, and P.~Zhang, ``{Joint
  Communication, Sensing, and Computation Enabled 6G Intelligent Machine
  System},'' \emph{IEEE Network}, vol.~35, no.~6, pp. 34--42, Nov. 2021.

\bibitem{liu2020joint}
F.~Liu, C.~Masouros, A.~Petropulu, H.~Griffiths, and L.~Hanzo, ``{Joint radar
  and communication design: Applications, state-of-the-art, and the road
  ahead},'' \emph{IEEE Transactions on Communications}, June 2020.

\bibitem{Chen2021CDOFDM}
X.~Chen, Z.~Feng, Z.~Wei, P.~Zhang, and X.~Yuan, ``{Code-Division OFDM Joint
  Communication and Sensing System for 6G Machine-Type Communication},''
  \emph{IEEE Internet of Things Journal}, vol.~8, no.~15, pp. 12\,093--12\,105,
  Feb. 2021.

\bibitem{ZhangOverviewJCS}
J.~A. Zhang, F.~Liu, C.~Masouros, R.~W. Heath, Z.~Feng, L.~Zheng, and
  A.~Petropulu, ``An overview of signal processing techniques for joint
  communication and radar sensing,'' \emph{IEEE Journal of Selected Topics in
  Signal Processing}, vol.~15, no.~6, pp. 1295--1315, Sept. 2021.

\bibitem{Yuan2021}
X.~Yuan, Z.~Feng, J.~A. Zhang, W.~Ni, R.~P. Liu, Z.~Wei, and C.~Xu,
  ``{Spatio-Temporal Power Optimization for MIMO Joint Communication and Radio
  Sensing Systems With Training Overhead},'' \emph{IEEE Transactions on
  Vehicular Technology}, vol.~70, no.~1, pp. 514--528, Jan. 2021.

\bibitem{Zhang2022ISAC}
J.~A. Zhang, K.~Wu, X.~Huang, Y.~J. Guo, D.~Zhang, and R.~W. Heath,
  ``Integration of radar sensing into communications with asynchronous
  transceivers,'' \emph{IEEE Communications Magazine}, pp. 1--7, Aug. 2022.

\bibitem{IBFDJCR}
S.~A. Hassani, B.~van Liempd, A.~Bourdoux, F.~Horlin, and S.~Pollin, ``Joint
  in-band full-duplex communication and radar processing,'' \emph{IEEE Systems
  Journal}, pp. 1--9, July 2021.

\bibitem{2021Girimlocalization}
G.~Kwon, A.~Conti, H.~Park, and M.~Z. Win, ``Joint communication and
  localization in millimeter wave networks,'' \emph{IEEE Journal of Selected
  Topics in Signal Processing}, vol.~15, no.~6, pp. 1439--1454, 2021.

\bibitem{Chen2023KF}
X.~Chen, Z.~Feng, J.~Andrew~Zhang, Z.~Wei, X.~Yuan, and P.~Zhang,
  ``Sensing-aided uplink channel estimation for joint communication and
  sensing,'' \emph{IEEE Wireless Communications Letters}, vol.~12, no.~3, pp.
  441--445, March 2023.

\bibitem{2019YuanLocalization}
W.~Yuan, N.~Wu, B.~Etzlinger, Y.~Li, C.~Yan, and L.~Hanzo,
  ``Expectation–maximization-based passive localization relying on
  asynchronous receivers: Centralized versus distributed implementations,''
  \emph{IEEE Transactions on Communications}, vol.~67, no.~1, pp. 668--681,
  Jan. 2019.

\bibitem{Qian2018Widar2}
K.~Qian, C.~Wu, Y.~Zhang, G.~Zhang, Z.~Yang, and Y.~Liu, ``Widar2.0: Passive
  human tracking with a single wi-fi link,'' \emph{Proceedings of the 16th
  Annual International Conference on Mobile Systems, Applications, and
  Services}, p. 350–361, 2018.

\bibitem{Nizhitong2021}
Z.~Ni, J.~A. Zhang, X.~Huang, K.~Yang, and J.~Yuan, ``Uplink sensing in
  perceptive mobile networks with asynchronous transceivers,'' \emph{IEEE
  Transactions on Signal Processing}, vol.~69, pp. 1287--1300, Feb. 2021.

\bibitem{ZhangDaqing2019}
Y.~Zeng, D.~Wu, J.~Xiong, E.~Yi, R.~Gao, and D.~Zhang, ``{FarSense: Pushing the
  Range Limit of WiFi-Based Respiration Sensing with CSI Ratio of Two
  Antennas},'' \emph{Proc. ACM Interact. Mob. Wearable Ubiquitous Technol.},
  vol.~3, no.~3, Sept. 2019.

\bibitem{ZhangDaqing2020}
Y.~Zeng, D.~Wu, J.~Xiong, J.~Liu, Z.~Liu, and D.~Zhang, ``Multisense: Enabling
  multi-person respiration sensing with commodity wifi,'' \emph{Proc. ACM
  Interact. Mob. Wearable Ubiquitous Technol.}, vol.~4, no.~3, Sept. 2020.

\bibitem{Li2022CSIsensing}
X.~Li, J.~Andrew~Zhang, K.~Wu, Y.~Cui, and X.~Jing, ``{CSI-Ratio-based Doppler
  Frequency Estimation in Integrated Sensing and Communications},'' \emph{IEEE
  Sensors Journal}, vol.~22, no.~21, pp. 20\,886--20\,895, Sept. 2022.

\bibitem{Sturm2011Waveform}
C.~Sturm and W.~Wiesbeck, ``{Waveform design and signal processing aspects for
  fusion of wireless communications and radar sensing},'' \emph{Proceedings of
  the IEEE}, vol.~99, no.~7, pp. 1236--1259, May 2011.

\bibitem{Zhang2019JCRS}
J.~A. {Zhang}, X.~{Huang}, Y.~J. {Guo}, J.~{Yuan}, and R.~W. {Heath},
  ``{Multibeam for joint communication and radar sensing using steerable analog
  antenna arrays},'' \emph{IEEE Transactions on Vehicular Technology}, vol.~68,
  no.~1, pp. 671--685, Jan. 2019.

\bibitem{2023XuJCAS}
X.~Chen, Z.~Feng, Z.~Wei, X.~Yuan, P.~Zhang, J.~Andrew~Zhang, and H.~Yang,
  ``Multiple signal classification based joint communication and sensing
  system,'' \emph{IEEE Transactions on Wireless Communications}, pp. 1--1, Feb.
  2023.

\bibitem{2023ChenKalman}
X.~Chen, Z.~Feng, J.~Andrew~Zhang, Z.~Wei, X.~Yuan, and P.~Zhang,
  ``Sensing-aided uplink channel estimation for joint communication and
  sensing,'' \emph{IEEE Wireless Communications Letters}, vol.~12, no.~3, pp.
  441--445, March 2023.

\bibitem{Chen2020}
X.~{Chen}, Z.~{Feng}, Z.~{Wei}, F.~{Gao}, and X.~{Yuan}, ``{Performance of
  Joint Sensing-Communication Cooperative Sensing UAV Network},'' \emph{IEEE
  Transactions on Vehicular Technology}, vol.~69, no.~12, pp. 15\,545--15\,556,
  Dec. 2020.

\bibitem{Rodger2014principles}
W.~H.~T. Rodger E.~Ziemer, \emph{Principles of Communications}, 7th~ed.\hskip
  1em plus 0.5em minus 0.4em\relax Wiley, 2014.

\bibitem{2010MIMO}
Y.~S. Cho, J.~Kim, W.~Y. Yang, and C.~G. Kang, \emph{MIMO-OFDM Wireless
  Communications with MATLAB}.\hskip 1em plus 0.5em minus 0.4em\relax Wiley
  Publishing, 2010.

\bibitem{HAARDT2014651}
M.~Haardt, M.~Pesavento, F.~Roemer, and M.~{Nabil El Korso}, ``Chapter 15 -
  subspace methods and exploitation of special array structures,'' in
  \emph{Academic Press Library in Signal Processing: Volume 3}, A.~M. Zoubir,
  M.~Viberg, R.~Chellappa, and S.~Theodoridis, Eds.\hskip 1em plus 0.5em minus
  0.4em\relax Elsevier, 2014, vol.~3, pp. 651--717.

\bibitem{Moor1993}
B.~De~Moor, ``The singular value decomposition and long and short spaces of
  noisy matrices,'' \emph{IEEE Transactions on Signal Processing}, vol.~41,
  no.~9, pp. 2826--2838, Sept. 1993.

\end{thebibliography}
}
\vspace{-10 mm}
\ifCLASSOPTIONcaptionsoff
  \newpage
\fi

\end{document}